\documentclass
[aps,prb,twocolumn,floatfix,english,showpacs,10pt,superscriptaddress]{revtex4-2}%
\usepackage[utf8]{inputenc}
\usepackage{amsmath}
\usepackage{physics}
\usepackage{braket}
\usepackage{ulem}
\usepackage{amssymb}
\usepackage{graphicx}
\usepackage{xcolor}
\usepackage[most]{tcolorbox}
\usepackage[unicode=true, breaklinks=false, pdfborder={0 0 1}, backref=false, colorlinks=true, linkcolor=blue, urlcolor=blue, citecolor=blue]{hyperref}
\setcitestyle{numbers,square}

\begin{document}

\title{Persistent nodal magnon-photon polariton in ferromagnetic heterostructures}

\author{Zhuolun Qiu}
\thanks{These authors contributed equally to this work.}
\affiliation{School of Physics, Huazhong University of Science and Technology, Wuhan 430074, China}

\author{Xi-Han Zhou}
\thanks{These authors contributed equally to this work.}
\affiliation{School of Physics, Huazhong University of Science and Technology, Wuhan 430074, China}

\author{Hanchen Wang}
\affiliation{Department of Materials, ETH Zurich, Zurich 8093, Switzerland}

\author{Guang Yang}
\affiliation{School of Integrated Circuit Science and Engineering, Beihang University, Beijing 100191, China}

\author{Tao Yu}
\email{taoyuphy@hust.edu.cn}
\affiliation{School of Physics, Huazhong University of Science and Technology, Wuhan 430074, China}
\date{\today }

\begin{abstract}
 Exceptional points with coalescence of eigenvalues and eigenvectors are spectral singularities in the parameter space, achieving which often needs fine-tuning of parameters in quantum systems.  We predict a \textit{persistent} realization of nodal magnon-photon polariton, \textit{i.e.}, a polariton of long wavelength without any gap splitting in a thin ferromagnetic insulator film sandwiched by two normal metals, which persistently exists when the ferromagnet is sufficiently thick $\sim 100$~nm due to the joint effect of dissipation and dissipative coupling. We perform the model calculation \textit{beyond the perturbation theory} using a classical approach, develop a quantum scheme able to account for the Ohmic dissipation, and find ultrastrong coupling with coupling strength comparable to the bare magnon frequency. Via revealing a simple conversion relation we extend this formalism to superconductors and predict the gap opened by the ultrastrong coupling strongly depends on the direction of polariton propagation.  Our findings may help search for robust non-Hermitian topological phases in magnonic and spintronic devices.

\end{abstract}

\maketitle

\section{Introduction}

Ferromagnetic heterostructures composed of ferromagnets (FMs) and normal metals (NMs) are common devices intensively explored and exploited in spintronics over decades, tracing back to the discovery of giant magnetoresistance~\cite{Nobel_spin_waves}. In these heterostructures, NM$|$FM, FM$|$NM$|$FM, or NM$|$FM$|$NM could be basic unit cells. Due to the Ohmic dissipation in NMs, these heterostructures are intrinsically dissipative and non-Hermitian, thus providing platforms for searching for useful non-Hermitian topological phases with magnetism~\cite{Flebus_review,yu2024nonhermitian,wang2020dissipative}. For example,  Liu
\textit{et al.} experimentally observed the exceptional points (EPs) of magnons, a state with both eigenstates and eigenvalues coalescing at specific parameters, in FM$|$NM$|$FM heterostructure in which the NM layer mediates a RKKY exchange interaction~\cite{liu2019observation}. This realization needs fine-tuning of parameters~\cite{liu2019observation,Xia,Yan}, including the gain, loss, and RKKY exchange interaction, which is device-dependent and, thereby, difficult to achieve universally.

It has been revealed that the NM affects the excitations of magnetization in FMs, i.e., spin waves or magnons as their quanta. They act as promising information carriers to transport spin information and perform logic operations in ``magnonics"~\cite{Chumak2015,Kruglyak2010,Cornelissen2015,Lebrun2018,cavity_magnonics,yu2024nonhermitian,roadmap2024}. The dipolar fields emitted by spin waves drive the diamagnetic eddy currents in the NMs, which are dissipated according to Ohm’s Law and thereby cause an additional dissipation channel of magnons~\cite{Imaging_spin-wave_Bertelli,Excitation_Pincus,Kostylev2009, Schoen2015, Flovik2015, Li2016, Kostylev2016, Rao2017, Balaji2017, Nikitin2018, Gladii2019, Serha2022, Bunyaev2020}.  Moreover, the chirality of the stray field translates into non-local chiral damping, i.e., it only affects spin waves that obey the right-hand rule, rendering non-Hermitian skin effect~\cite{bergholtz2021exceptional,elganainy2018nonhermitian,miri2019exceptional,yu2024nonhermitian} and unidirectional transmission across potential barriers~\cite{Xiyin}. However, despite being useful for spintronic and magnonic devices, the studies addressing collective modes in the NM$|$FM$|$NM systems are still limited, in which the magnon-photon interaction could be much enhanced due to the ``cavity effect".

Different functionalities appear when the NMs become superconducting at low temperatures since the Ohmic dissipation disappears. The induced eddy currents in superconductors (SCs) also generate Oersted magnetic fields that blue-shift the magnon frequencies~\cite{CPL_exp,PRA_exp,PRA_exp2,Gient_demagnetization,zhou2023gating,Silaev,shift_theory}. Superconducting gates, therefore, act as non-dissipative repulsive potential barriers for propagating spin waves~\cite{Volkov_junction,Yu_gating,Borst,superconductor_gate,superconductor_gate_1,superconductor_gate_2,superconductor_gate_3}. Recently, ultrastrong coupling~\cite{Nori_ultrastrong} between magnon and so-called Swihart photon modes of SCs~\cite{swihart} was reported in SC$|$FM$|$SC Josephson junctions~\cite{silaev2023ultrastrong,golovchanskiy2021approaching,golovchanskiy2021ultrastrong}. Recent experiments~\cite{golovchanskiy2021approaching,golovchanskiy2021ultrastrong} addressed a large anticrossing between Kittel magnon and Swihart photon in SC-FM multilayers, thereby inspiring the on-chip realization of ultrastrong coupling with magnons.  Silaev~\cite{silaev2023ultrastrong} predicted ultrastrong coupling between propagating magnon with Swihart photon in SC$|$FM$|$SC heterostructures, useful for quantum squeezed vacuum states, but focusing on the bulk volume configuration with wave vectors aligned with the saturation magnetization. Ultrastrong coupling comparable to the bare magnon frequency goes beyond the conventional devices with a millimeter-sized magnetic sphere loaded into a metallic microwave cavity~\cite{cavity_magnonics,wang2020dissipative,harder2021coherent,wang2024enhancement,zhang2016cavity,zhang2014strongly,huebl2013high, Ding2018Anisotropic}, in which case similar coupling strength requires much larger magnet/cavity volume ratio~\cite{hou2019strong,li2019strong,guo2023strong,Yu_gating,mckenzie-sell2019low}.

\begin{figure}[htp]
\includegraphics[width=89mm]{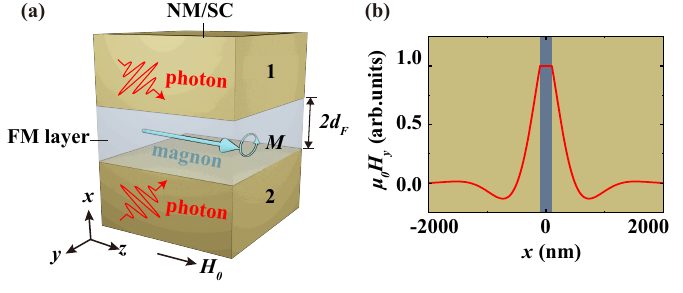}
\caption{(a) Ferromagnetic heterostructures consisting of two outer layers of normal metals (or superconductors) and a middle ferromagnetic insulator, biased by an in-plane magnetic field $H_0\hat{\bf z}$. (b) Confinement of stray magnetic field $\mu_0 H_y$ across the heterostructure thickness direction when the thickness of a middle layer is 200~nm, and the thickness of surrounding conductors is several micrometers.}
\label{model}
\end{figure}

In this work, we investigate collective modes in the ferromagnetic insulator (FI) sandwiched by two NMs, as illustrated in Fig.~\ref{model}(a). In such NM$|$FI$|$NM heterostructure, we predict the coupling between the magnon and photon is ultrastrong with the coupling strength comparable to the bare magnon frequency due to the strong confinement of microwave magnetic field across the heterostructure [Fig.~\ref{model}(b)]. We develop a non-Hermitian quantization scheme with non-Hermitian operators that allows for accounting the Ohmic dissipation, different from the scheme for the LRC circuit (e.g., Ref.~\cite{quantization}). The quantum formalism shows analytically how the photon and magnon couple dissipatively. By going beyond the perturbation theory, we find unexpectedly that the eddy-current-induced damping favors the EPs with the coalescence of both eigenstates and eigenvalues at a specific wave vector~\cite{heiss2012physics,bender2007making}. Surprisingly, such EPs in this heterostructure persistently exist in the wave-vector space when the thickness of the FI is thicker than a ``critical" one. Recently, EPs have been experimentally demonstrated in magnonic systems~\cite{zhang2019experimental, liu2019observation,wang2023floquet,harder2017topological,zhang2019higher,zhang2017observation,wang2024enhancement}, but the realization of such singularity lies in the perturbation theory and needs fine-tuning of the parameters. Finally, we extend the formalism to the SC$|$FI$|$SC Josephson junction by revealing a close connection between the two systems: the formalisms with NMs connect with those with SCs by simply replacing the skin depth $\delta$ with London's penetration depth $\lambda$. We find strong anisotropy in the coupling: the coupling is ultrastrong in the bulk volume configuration (the wave vector ${\bf k}\parallel \hat{\bf z}$ in Fig.~\ref{model}) but vanishes in the  Damon-Eshbach case (${\bf k}\perp \hat{\bf z}$) due to the chirality of magnons. 

This article is organized as follows. In Sec.~\ref{Sec_II}, we introduce the model and the basic equations for solving the collective modes in the ferromagnetic heterostructures. We focus on the NM$|$FI$|$NM heterostructure in Sec.~\ref{Sec_III} and predict the robust non-Hermitian nodal polariton. Section~\ref{Sec_IV} shows the anisotropy in the magnon-photon ultrastrong coupling when the NMs become superconducting at low temperatures. We summarize our results and give an outlook in Sec.~\ref{Sec_V}.

\section{Model and formalism}
\label{Sec_II} 

We consider a FI of thickness $2d_F$ sandwiched by two thick NMs or SCs of thickness $\gg \lambda$, where $\lambda\sim O(100~{\rm nm})$ is the penetration depth of microwaves in the NMs or SCs. The interfacial exchange interaction that induces additional damping of magnons due to the spin pumping effect~\cite{tserkovnyak2002spin,tserkovnyak2002enhanced} can be suppressed by inserted thin insulators without affecting the long-range dipolar interaction. The in-plane magnetic field $H_0\hat{\bf z}$ along the $\hat{\bf z}$-direction biases the saturation magnetization ${\bf M}_{s}$, as shown in Fig.~\ref{model}(a).  For a thin ferromagnetic film, the magnetization fluctuation of long wavelength is nearly uniform across the film normal $\hat{\bf x}$-direction. In the linear response regime, we focus on the transverse fluctuation $M_{x(y)}({\bf r},t)={\cal M}_{x(y)} e^{i {\bf k}\cdot {\pmb \rho}-i\omega t}$ with amplitude ${\cal M}_{x(y)}$, wave vector ${\bf k}=k_y\hat{\bf y}+k_z\hat{\bf z}$, and frequency $\omega$, where ${\pmb \rho}=y\hat{\bf y}+z\hat{\bf z}$.  The transverse magnetization fluctuation radiates the electric and magnetic fields, captured by Maxwell's equations~\cite{Jackson}.

The electric field obeys Maxwell's equations~\cite{Jackson}
\begin{subequations}
\begin{align}
    &\nabla\times {\bf E}({\bf r}, t)=-\frac{\partial {\bf B}({\bf r}, t)}{\partial t},\label{maxwell_E}\\
    &\nabla \times {\bf H}({\bf r}, t)={\bf J}({\bf r}, t)+\epsilon_r\dfrac{\partial{\bf E}({\bf r}, t)}{\partial t},
    \label{maxwell_H}
\end{align}
\end{subequations}
where $\epsilon_r$ is the dielectric constant depending on the materials: $\epsilon_r=\epsilon_{\rm FI}$ in the FI, while in the SC and NM, $\epsilon_r=\epsilon_0$ is the dielectric constant in the vacuum. 
${\bf J}({\bf r}, t)$ is the electric current.  In the FI, ${\bf J}({\bf r}, t)=0$, so taking the curl of Eq.~(\ref{maxwell_E}) and substituting it into (\ref{maxwell_H}) yields 
\begin{align}
\text{in FI}: ~~\nabla^2 \mathbf{E}({\bf r}, t)+k^2_0 \mathbf{E}({\bf r}, t)
=-i\omega \mu_0  \nabla\times {\bf M}({\bf r},t),
\label{electric_field}
\end{align} 
where $\mu_0$ is the  vacuum permeability and $k_0\equiv \omega\sqrt{\mu_0\epsilon_{\rm FI}}$. Equation~\eqref{electric_field} implies that the ``magnetization current" $\nabla\times{\bf M}$ radiates the electric field.
In the NM, the electric field drives the electric current according to Ohm's Law ${\bf J}({\bf r}, t)=\sigma_c {\bf E}({\bf r}, t)$,
where $\sigma_c$ is the electric conductivity.
The electric field in the NM is governed by 
\begin{align}
\text{in NM}: ~~~ \nabla^2 \mathbf{E}({\bf r}, t)+k_n^2\mathbf{E}({\bf r}, t)=0,
\end{align}
 where $k_{n}=\sqrt{\omega^2\mu_0\epsilon_0+i\omega\mu_0\sigma_c}$.

 When the NM enters into the superconducting states at low temperatures, the electric field, governed by London's equation $\partial {\bf J}({\bf r},t)/\partial t={\bf E}({\bf r},t)/(\mu_0\lambda^2)$, drives the ``supercurrent" ${\bf J}({\bf r},t)=\sigma_s{\bf E}({\bf r},t)$ that is out of phase of the electric field since the ``conductivity" $\sigma_s=i/(\omega\mu_0\lambda^2)$ of SCs is purely imaginary when far below the superconducting transition temperature $T_c$, such that the electric field inside the SCs obeys
 \begin{align}
     \text{in SC}: ~~~  \nabla^2 \mathbf{E}({\bf r}, t)+k_s^2\mathbf{E}({\bf r}, t)=0,
 \end{align}
 where $k_s=\sqrt{\omega^2\mu_0\epsilon_0-1/\lambda^2}$ depends on the London's penetration depth $\lambda$.

With the spin waves $M_{x(y)}={\cal M}_{x(y)}e^{i{\bf k}\cdot {\pmb \rho}-i\omega t}$, we assume the formal solution 
 ${\bf E}({\bf r},t)={\tilde {\bf E}}(x)e^{i{\bf k}\cdot {\pmb \rho}-i\omega t}$, in which the amplitudes 
 \begin{subequations}
\begin{align}
    &\text{in FI}:~~{\tilde {\bf E}}(x)={\bf U} +{\pmb{\cal E}}_{0}e^{i {\cal A}_k x}+{\pmb{\cal E}}_{0}'e^{-i {\cal A}_k x}\label{sol_F},\\
    &\text{in NM/SC~``1"}~(x>d_{F}):~~{\tilde {\bf E}}(x)={\pmb{\cal E}}_{1}e^{i {\cal B}_k x},\label{sol_F1}\\
  &\text{in NM/SC~``2"}~(x<-d_{F}):~~{\tilde {\bf E}}(x)={\pmb{\cal E}}_{2}e^{-i {\cal B}_k x},\label{sol_F2}
 \end{align}
\end{subequations}
where ${\cal A}_k=\sqrt{k_0^2-k^2}$ and in the NM (SC) ${\cal B}_k=\sqrt{k_{n(s)}^2-k^2}$. In Eq.~(\ref{sol_F}), ${\bf U}={\omega\mu_0}(-k_z {\cal M}_{y} {\hat{\bf x}}+k_z{\cal M}_{x}{\hat{\bf y}} -k_y {\cal M}_{x}{\hat{\bf z}})/{\cal A}_k^2$ is the special solution of Eq.~(\ref{electric_field}).

The oscillating electric field converts to the magnetic field according to Faraday's Law, i.e., $i\omega\mu_0[{\bf H}({\bf r},t)+{\bf M}({\bf r},t)]=\nabla\times {\bf E}({\bf r},t)$, leading to 
\begin{subequations}
     \begin{align}
     \label{magnetic_field_y}
     &H_y=1/(i \omega \mu_0)(\partial_z E_x-\partial_x E_z)-M_y,\\
     \label{magnetic_field_z}
     &H_z=1/(i \omega\mu_0)(\partial_x E_y-\partial_y E_x),\\
     &H_x=1/(i \omega\mu_0)(\partial_y E_z-\partial_z E_y)-M_x.
     \label{magnetic_field_x}
 \end{align}
 \label{magnetic_field}
\end{subequations}

The coefficients in ${\tilde {\bf E}}(x)$ are then solved by the boundary conditions of the electric and magnetic fields at material interfaces. The in-plane component ${\bf E}_{\parallel}$ is continuous at the interfaces. On the other hand, for the ``good" NM conductors and SCs, the characteristic relaxation time of electrons $\tau\ll1/\omega$, implying no charge accumulation is built inside the conductors and hence $\nabla\cdot {\bf E}({\bf r},t)=0$ therein. Then by the continuity of ${\bf E}_{\parallel}$,  $\partial_x{E}_x({\bf r},t)$ is continuous at the interfaces~\cite{swihart}.  For the magnetic field, at the interfaces ${\bf B}_\perp$ and ${\bf H}_\parallel$ are continuous~\cite{Jackson}.

To solve ${E}_x({\bf r},t)$, we need another boundary condition, however. The $x$-component of Eq.~(\ref{maxwell_H}) obeys
$i\omega\mu_0(ik_yH_z-ik_zH_y)=\left(i\omega\mu_0\sigma_{c(s)}+\omega^2\mu_0\epsilon_r\right){E}_x$.
Then the continuity of $H_{y}$ and $H_z$ implies that the current  $(i\omega\mu_0\sigma_{c(s)}+\omega^2\mu_0\epsilon_r)E_x$ is continuous at the interfaces.
Accordingly, we find inside the FI   
\begin{align}
    {\cal E}_{0x}={\cal E}_{0x}'=\dfrac{R_kU_x}{(1-R_k)e^{i{\cal A}_k d_F}-(1+R_k)e^{-i A_k d_F}},
    \label{Ex}
\end{align}
where the factor $R_k=[({\cal A}_k^2+k^2){\cal B}_k]/[({\cal B}_k^2+k^2){\cal A}_k]$ depends on the surroundings.

For $E_y$, from the continuity of $E_y$ and $H_z$ at the interfaces, we find
\begin{align}
    {\cal E}_{0y}&=\dfrac{k_y\left(\tilde{E}_x|_{d_F^-}-\tilde{E}_x|_{d_F^+}\right)}{({\cal B}_k+{\cal A}_k)e^{-i {\cal A}_k d_F}-({\cal B}_k-{\cal A}_k)e^{i {\cal A}_k d_F}}\nonumber\\
    &-\dfrac{{\cal B}_k U_y}{({\cal B}_k+{\cal A}_k)e^{-i {\cal A}_k d_F}+({\cal B}_k-{\cal A}_k)e^{i {\cal A}_k d_F}},\nonumber\\
    {\cal E}_{0y}'&=-\dfrac{{\cal B}_k U_y}{({\cal B}_k+{\cal A}_k)e^{-i {\cal A}_k d_F}+({\cal B}_k-{\cal A}_k)e^{i {\cal A}_k d_F}}\nonumber\\
    &-\dfrac{k_y\left(\tilde{E}_x|_{d_F^-}-\tilde{E}_x|_{d_F^+}\right)}{({\cal B}_k+{\cal A}_k)e^{-i {\cal A}_k d_F}-({\cal B}_k-{\cal A}_k)e^{i {\cal A}_k d_F}},
    \label{Ey}
\end{align}
in which,  according to Eq.~\eqref{sol_F} and the continuity of $H_y$ and $H_z$, $\tilde{E}_x|_{d_F^{-}}=U_x+{\cal E}_{0x}e^{i {\cal A}_k d_F}+{\cal E}'_{0x}e^{-i {\cal A}_k d_F}$ and $\tilde{E}_x|_{d_F^{+}}=(i\omega\mu_0\sigma_{c(s)}+\omega^2\mu_0\epsilon_0) /(\omega^2\mu_0\epsilon_{\rm FI})\tilde{E}_x|_{d_F^{-}}$. These components depend explicitly on $k_y$ and are thereby anisotropic.

Similarly, we solve the amplitudes in $E_z$ as
\begin{align}
    {\cal E}_{0z}&=-\dfrac{1}{k_z}\left[\dfrac{{\cal A}_k R_k U_x}{(1-R_k)e^{i{\cal A}_k d_F}-(1+R_k)e^{-i {\cal A}_k d_F}}\right.\nonumber\\
    &+\dfrac{k_y^2\left(\tilde{E}_x|_{d_F^-}-\tilde{E}_x|_{d_F^+}\right)}{({\cal B}_k+{\cal A}_k)e^{-i {\cal A}_k d_F}-({\cal B}_k-{\cal A}_k)e^{i {\cal A}_k d_F}}\nonumber\\
    &\left.-\dfrac{k_y{\cal B}_k U_y}{({\cal B}_k+{\cal A}_k)e^{-i {\cal A}_k d_F}+({\cal B}_k-{\cal A}_k)e^{i {\cal A}_k d_F}}\right],\nonumber\\
    {\cal E}_{0z}'&=\dfrac{1}{k_z}\left[
\dfrac{{\cal A}_k R_k U_x}{(1-R_k)e^{i{\cal A}_k d_F}-(1+R_k)e^{-i {\cal A}_k d_F}}\right.\nonumber\\
&+\dfrac{k_y^2\left(\tilde{E}_x|_{d_F^-}
    -\tilde{E}_x|_{d_F^+}\right)}{({\cal B}_k+{\cal A}_k)e^{-i {\cal A}_k d_F}-({\cal B}_k-{\cal A}_k)e^{i {\cal A}_k d_F}}\nonumber\\
    &+\left.\dfrac{k_y{\cal B}_k U_y}{({\cal B}_k+{\cal A}_k)e^{-i {\cal A}_k d_F}+({\cal B}_k-{\cal A}_k)e^{i {\cal A}_k d_F}}\right].
    \label{Ez}
\end{align}

With the solution of electric fields Eqs.~\eqref{Ex}, \eqref{Ey}, and \eqref{Ez}, we find the magnetic field via Eq.~\eqref{magnetic_field}.
For the thin FI, the radiated magnetic field has a long wavelength and is, therefore, nearly uniform across the thickness. We thereby take its value at $x=0$ as the averaged one, i.e., ${H}_{x,y}={H}_{x,y}|_{x=0}={\cal H}_{x,y}e^{i{\bf k}\cdot{\pmb \rho}-i\omega t}$, where ${\cal H}_{x,y}$ are the amplitudes. According to Eqs.~(\ref{magnetic_field_y}) and (\ref{magnetic_field_x}), these amplitudes
\begin{align}
    {\cal H}_{y}&=-{\cal M}_{y}+({k_z}/{\omega\mu_0})(U_x+2{\cal E}_{0x})\nonumber\\
    &+\frac{{2\cal A}_k}{\omega\mu_0k_z}\left[\frac{{\cal A}_k R_k U_x}{(1-R_k)e^{i{\cal A}_kd_F}-(1+R_k)e^{-i{\cal A}_kd_F}}\right.\nonumber\\
    &\left.+\frac{k_y^2(1-{k_0^2}/{k_{n(s)}^2})(U_x+{\cal E}_{0x}(e^{i{\cal A}_kd_F}+e^{-i{\cal A}_kd_F}))}{({\cal B}_k+{\cal A}_k)e^{-i{\cal A}_kd_F}-({\cal B}_k-{\cal A}_k)e^{i{\cal A}_kd_F}}\right],\nonumber\\
    {\cal H}_{x}&=-{\cal M}_{x}+({k_y}/{\omega\mu_0})\nonumber\\
    &\times\left[U_z+\frac{2k_y{\cal B}_kU_y}{k_z(({\cal B}_k+{\cal A}_k)e^{-i{\cal A}_kd_F}+({\cal B}_k-{\cal A}_k)e^{i{\cal A}_kd_F})}\right]\nonumber\\
    &-\frac{k_z}{\omega \mu_0}\left[U_y-\frac{2{\cal B}_kU_y}{({\cal B}_k+{\cal A}_k)e^{-i{\cal A}_kd_F}+({\cal B}_k-{\cal A}_k)e^{i{\cal A}_kd_F}}\right],
    \label{amplitudes_magnetic_field}
\end{align}
in which the amplitudes of magnetization ${\cal M}_{x,y}$ is solved by combining the Landau-Lifshitz-Gilbert equation.

\section{NM$|$FI$|$NM heterostructure}
\label{Sec_III} 

\subsection{Classical approach}

We can significantly simplify the amplitudes of the magnetic field  \eqref{amplitudes_magnetic_field} when the frequency $\omega$ of the collective modes in the NM$|$FI$|$NM heterostructure is low, and their wavelength $2\pi/k$ is long.
When $k<10^4$ rad/m and $\omega\ll \omega_c$, where $\omega_c\equiv{\sigma_{c}}/{\epsilon}_0$ is the plasma frequency in the NMs~\cite{Jackson}, ${\cal B}_k=\sqrt{\omega^2\mu_0\epsilon_0+i\omega\mu_0\sigma_{c}-k^2}\approx\sqrt{i\omega\mu_0\sigma_{c}}= i/\delta$, which is justified by $|i\omega\mu_0\sigma_{c}|=3.7\times 10^{12}~{\rm m}^{-2} \gg |\omega^2\mu_0\epsilon_0-k^2|= 1.0\times 10^8~{\rm m}^{-2}$ when the conductivity $\sigma_c\sim6\times 10^{7}~\Omega^{-1}\cdot {\rm m}^{-1}$, where $\delta=(1+i)/\sqrt{2\omega\mu_0\sigma_{c}}$ is the penetration depth of microwaves in NMs that contains both the real and imaginary components, accounting for the oscillatory decay of microwaves in metals. For typical conductivity of copper $\sigma_{c}=5.96\times 10^7$~$\Omega^{-1}\cdot {\rm m}^{-1}$, the approximation is suitable as long as  $\omega\ll \omega_c\sim 10^{9}$~GHz such that no charge accumulation is built up inside the conductors. Furthermore, when $k<10^4$ rad/m and $d_F\lesssim 1~\mu$m, $|{\cal A}_k|=\left|\sqrt{k_0^2-k^2}\right|<k$ and $|{\cal A}_kd_F|<k d_F\ll 1$. For example, for thin ferromagnetic films of thickness $d_F\sim 100$~nm, typical frequency $\omega\sim 50$~GHz, and wave vector $k\sim 10^4$~${\rm m}^{-1}$, which is comparable to the wavelength of light $k_0=\omega\sqrt{\mu_0 \epsilon_{\rm FI}}$, the expansion is guaranteed by $|{\cal A}_k| d_F=|\sqrt{k_0^2-k^2}| d_F\approx0.001\ll1 $.
Therefore, for thin FI layers, we are allowed to expand $e^{\pm i{\cal A}_kd_F}$  to the leading order $1\pm i{\cal A}_kd_F$. 
The amplitudes of magnetic fields \eqref{amplitudes_magnetic_field} are then reduced to
\begin{align}
    &{\cal H}_{x}=-{\cal M}_{x} \dfrac{1-k_0^2d_F\delta}{1-{\cal A}_k^2d_F\delta}\approx -{\cal M}_{x},\nonumber\\
   &{\cal H}_{y}=-{\cal M}_{y} \dfrac{d_F}{k_0^2\delta+{\cal A}_k^2d_F}\left(k_0^2-\frac{d_F}{\delta+d_F}k_y^2\right),
   \label{C}
\end{align}
such that $H_x$ is out of phase $\pi$ to $M_x$, while $H_y$ contains the component out of phase $\pi/2$ to $M_y$ due to the imaginary component of $\delta$.

By substitution of Eq.~\eqref{C} into the (linearized) Landau-Lifshitz-Gilbert equation
    \begin{align}
    -i\omega M_x+\mu_0\gamma H_0M_y-\mu_0\gamma M_s{H}_y -i\alpha_G\omega M_y&=0,\nonumber\\
    \mu_0\gamma H_0 M_x-\mu_0\gamma M_s{H}_x+i\omega M_y-i\alpha_G\omega M_x&=0,
    \label{LLG}
    \end{align}
    where $\alpha_G$ describes phenomenologically the intrinsic Gilbert damping of the ferromagnetic material, which also accounts for the relations between amplitudes after the substitution of $M_{x,y}={\cal M}_{x,y}e^{i{\bf k}\cdot{\pmb \rho}-i\omega t}$ and $H_{x,y}={\cal H}_{x,y}e^{i{\bf k}\cdot{\pmb \rho}-i\omega t}$,
we arrive at the coupled equation
 \begin{widetext} 
\begin{equation}
    \left(\begin{array}{cccc}
         0 & -\mu_0\gamma M_{s} & -i\omega & \mu_0\gamma H_0-i \alpha_G\omega \\
         -\mu_0\gamma M_{s} & 0 & \mu_0\gamma H_0-i \alpha_G\omega & i\omega \\
         1 & 0 & 1 & 0 \\
         0 & \alpha_k & 0 & \beta_{\bf k} \\
    \end{array}\right)\\
    \left(\begin{array}{c}
        H_x\\H_y\\M_x\\M_y
    \end{array}\right)=0,\nonumber
\end{equation}
where $\alpha_k=k_0^2(\delta+d_F)-k^2d_F$ and $\beta_{\bf k}=k_0^2d_F-{d^2_F}{/(\delta+d_F})k_y^2$ are both complex numbers. It leads to the characteristic equation for solving the frequency of the collective polariton modes  
\begin{equation}
    k^2+K_{\omega}\frac{d_F}{d_F +\delta}k_y^2=k_0^2\left(1+K_\omega+\frac{\delta}{d_F}\right),
    \label{nodal_magnon_polariton}
\end{equation}
in which 
\begin{align}
    K_\omega&=\frac{\mu_0\gamma M_{s}[\mu_0\gamma(M_{s}+H_0)-i\alpha_G \omega]}{\mu_0\gamma H_0[\mu_0\gamma(M_{s}+H_0)-2i\alpha_G \omega]-\omega[\omega(1+\alpha_G^2)+i\alpha_G\mu_0\gamma M_{s}]}.
    \label{K_w}
\end{align}
\end{widetext}

We shall resort to the numerical calculation to solve Eq.~\eqref{nodal_magnon_polariton}.
Since the penetration depth $\delta$ of microwaves in the NMs is a complex quantity, the solution of \eqref{nodal_magnon_polariton} for the collective-mode dispersion is also a complex number. The complex component of $\omega$ represents the inverse lifetime of collective modes.  In our system, the dissipation is introduced by the Ohmic dissipation of the eddy current in the NM layers and the intrinsic Gilbert damping of the ferromagnetic materials.

\subsection{Quantum scheme with Ohmic dissipation}

In the NM$|$FI$|$NM heterostructure, the photon modes are well confined between two NM layers in the scale of $\sim 100$~nm, acting as an ultrathin microwave waveguide. Magnon modes in the FI are also renormalized since the stray magnetic field is changed due to the back-and-forth reflection by the NMs according to Eq.~\eqref{C}. 
The interaction between magnon and photon in the ferromagnetic heterostructure can be conveniently understood using the quantum approach. However, the Ohmic dissipation in the  NM$|$FI$|$NM heterostructure presents a challenge that requires a different quantum scheme from Hermitian conventions.

In the long wavelength limit, the eigenfrequency of the magnon modes is captured by Kittel's formula~\cite{Kittel_mode}  
\begin{equation}
\omega_N=\mu_0\gamma\sqrt{(H_0+N_{xx}M_{s})(H_0+N_{yy}M_{s})},
\label{kittel_NM}
\end{equation}
where $N_{xx}$ and $N_{yy}$ are the ``demagnetization factors". Without the NMs, $N_{xx}=1$ and $N_{yy}=0$ are real numbers for the magnetic films. Nevertheless, the NMs reflect and dissipate the magnetic stray field, rendering the demagnetization factors to be complex numbers. According to Eq.~\eqref{C}, when $k\rightarrow0$ (refer to Appendix~\ref{demagnetization_factor} for explicit derivation of the renormalized demagnetization factor)
\begin{equation}
    N_{xx}=1,~~~N_{yy}={d_F}/({\delta+d_F}).
\end{equation}
So $N_{xx}$ is a constant, while $N_{yy}$ depends on the collective-mode frequency $\omega$ via $\delta$. By substitution into Kittel's formula \eqref{kittel_NM}, the eigenfrequency of the magnon mode in the NM$|$FI$|$NM is strongly renormalized to be  
 \begin{equation}
\omega_N=\mu_0\gamma\sqrt{(H_0+M_s)\left(H_0+\frac{d_F}{\delta+d_F}M_s\right)},
 \end{equation}
 which is a complex number and cannot be described by the Gilbert phenomenology, different from that of the NM$|$FM bilayer~\cite{Xiyin}. Microwave absorption experiments should conveniently measure ferromagnetic resonance (FMR) with significant frequency shift and broadening.

Combining the Landau-Lifshitz-Gilbert equation \eqref{LLG} and the normalization condition $\int d\mathbf{r}\left({M}_x{M}_y^{*}-{M}_x^{*}{M}_y \right)=-i/2$ for the magnon modes~\cite{FMR_Walker,Spinwaveexcitations}, we find the normalized amplitudes
\begin{align}
    &{\cal M}_{x}=i\sqrt{\frac{\omega_N}{8d_F\mu_0\gamma(H_0+M_{s})}},\nonumber\\
    &{\cal M}_{y}=-\sqrt{\frac{\mu_0\gamma(H_0+M_{s})}{8d_F\omega_N}},
    \nonumber
\end{align}
with which we expand the magnetization operators as  \begin{align}
    \hat{M}_x&=-i\int\frac{d{\bf k}}{2\pi} \sqrt{\frac{\hbar M_s\omega_N}{4d_F\mu_0(H_0+M_s)}}\left(e^{i{\bf k}\cdot{\pmb \rho}}\hat{m}_{\bf k}-e^{-i{\bf k}\cdot{\pmb \rho}}\hat{m}_{\bf k}^{\dagger}\right),\nonumber\\
    \hat{M}_y&=\int\frac{d{\bf k}}{2\pi} \sqrt{\frac{\hbar\mu_0\gamma^2(H_0+M_s)M_s}{4d_F\omega_N}}\left(e^{i{\bf k}\cdot{\pmb \rho}}\hat{m}_{\bf k}+e^{-i{\bf k}\cdot{\pmb \rho}}\hat{m}_{\bf k}^{\dagger}\right),
    \label{magnetization_NM}
\end{align}
where  $\hat{m}_{\bf k}^{\dagger}$ and $\hat{m}_{\bf k}$ are the creation and annihilation operators of magnon. In this ansatz, $\hat{M}_{x,y}$ are no longer Hermitian operators since $\omega_N$ is not purely real, different from the conventional quantization scheme.

To derive the \textit{bare} photon modes in the heterostructure, we construct an alternative NM$|$non-magnetic-insulator(I)$|$NM  heterostructure, in which the thickness of the non-magnetic insulator is also set to be $2d_{F}$. The electric fields are governed by
\begin{align}
    \text{in I}: ~~~\nabla^2 \mathbf{E}({\bf r}, t)+k_0^2 \mathbf{E}({\bf r}, t)=0,\nonumber\\
    \text{in NMs}: ~~~\nabla^2 \mathbf{E}({\bf r}, t)+k_n^2 \mathbf{E}({\bf r}, t)=0.
    \label{field_NM}
\end{align}
Given the isotropy in the propagation $y$-$z$ plane, without losing generality, we consider the wave propagation along the $\hat{\bf z}$-direction. We assume the plane-wave solution ${\bf E}=\tilde{\bf E}(x)e^{ikz-i\omega t}$, where the amplitudes
\begin{align}
    &\text{In I},~~\tilde{\bf E}(x)=\pmb{ \cal E}_0e^{i{\cal A}_k x}+\pmb{ \cal E}'_0e^{-i{\cal A}_k x},\nonumber\\
    &\text{In NM}~(x>d_{F}),~~\tilde{\bf E}(x)=\pmb{ \cal E}_{1}e^{i{\cal B}_k x},\nonumber\\
    &\text{In NM}~(x<-d_{F}),~~\tilde{\bf E}(x)=\pmb{ \cal E}_{2}e^{-i{\cal B}_k x},
    \nonumber
\end{align}
in which ${\cal A}_k=\sqrt{k_0^2-k^2}$ and ${\cal B}_k=\sqrt{k_n^2-k^2}$. According to Faraday's Law,  the magnetic field $H_y=1/(i \omega \mu_0)(\partial_z E_x-\partial_x E_z)$. ${E}_z$ and $\partial_x E_x$ at the interfaces are continuous, leading to the secular equation for the components $\{{\cal E}_{0x},{\cal E}'_{0x}\}$:  
\begin{align}
    \left|
    \begin{matrix}
    (R_k-1)e^{i{\cal A}_k d_{F}} & (R_k+1)e^{-i{\cal A}_k d_{F}}  \\
    (R_k+1)e^{-i{\cal A}_k d_{F}} & (R_k-1)e^{i{\cal A}_k d_{F}} 
    \end{matrix}\right|=0,\nonumber
\end{align}
i.e., 
\begin{equation}
    i(R_{k}^2+1)\sin(2{\cal A}_kd_{F})=2R_{k}\cos(2{\cal A}_kd_{F}).
    \label{result_secular_equation}
\end{equation}
Since the wavelength $2\pi/|{\bf k}|$ is much larger than the insulator thickness $d_{F}$, $\sin(2{\cal A}_kd_{F})\approx2{\cal A}_kd_{F}$ and $\cos(2{\cal A}_kd_{F})\approx1$. Further,  when the frequency  $\lesssim 10$~GHz, $\omega\mu_0\sigma_c\gg\omega^2\mu_0\epsilon_0$ for the NMs, such that ${\cal B}_k\approx i/\delta$ and 
\begin{equation}
    R_k=\frac{({\cal A}_k^2+k^2){\cal B}_k}{{\cal A}_k({\cal B}_k^2+k^2)}\approx-i\frac{k_0^2}{{\cal A}_k}\delta\ll1
    .\nonumber
\end{equation}
Accordingly, Eq.~\eqref{result_secular_equation} is simplified as 
\begin{equation}
    k^2=\Omega_n^2\mu_0\epsilon_{\rm I}\left(1+{\delta}/{d_{F}}\right),
    \nonumber
\end{equation}
where $\epsilon_{\rm I}$ is the dielectric constant of the insulator, from which 
\begin{align}
\Omega_n=\sqrt{d_{F}/(d_{F}+\delta)\mu_0\epsilon_{\rm I}}|{\bf k}|
\label{photon_dispersion}
\end{align}
is the dispersion of photon modes in the NM$|$I$|$NM heterostructure, which is no longer purely real since $\delta$ contains the imaginary component. They correspond to the TM modes with $E_y=H_x=H_z=0$ and only have three components $\{E_z,E_x,H_y\}$.

 When $-d_F<x<d_F$, the electromagnetic fields derived from Eq.~\eqref{field_NM} read
\begin{align}
    \tilde{E}_{x,k}(x)&={\cal E}_{0x}e^{i{\cal A}_k x}+{\cal E}_{0x}'e^{-i{\cal A}_k x},\nonumber\\
    \tilde{E}_{z,k}(x)&=\frac{{\cal A}_k}{k}\left({\cal E}_{0x}e^{i{\cal A}_k x}-{\cal E}_{0x}'e^{-i{\cal A}_k x}\right),\nonumber\\
    \tilde{H}_{y,k}(x)&=\frac{{\omega \epsilon_{\rm I} }}{k}\left({\cal E}_{0x}e^{i{\cal A}_k x}+{\cal E}_{0x}'e^{-i{\cal A}_k x}\right),
    \nonumber
\end{align}
where ${\cal E}_{0x}'=(1-{R}_k)/(1+{R}_k)e^{2i{\cal A}_kd_F}{\cal E}_{0x}\approx{\cal E}_{0x}$, and  with the dispersion \eqref{photon_dispersion} of the photon mode 
\begin{align}
     {\cal A}_k&=\sqrt{k_0^2-k^2}=ik\sqrt{\frac{\delta}{d_F+\delta}},\nonumber\\
     {\cal B}_k&=\sqrt{k_n^2-k^2}=i\sqrt{\frac{1}{\delta^2}+\frac{\delta}{d_F+\delta}k^2}\approx \frac{i}{\delta}.\nonumber
\end{align}
Accordingly, the fields in the insulator are simplified as 
\begin{align}
    \tilde{E}_{x,k}(x)&\approx{\cal E}_{0x}(e^{i{\cal A}_k x}+e^{-i{\cal A}_k x})\approx 2{\cal E}_{0x},\nonumber\\
    \tilde{E}_{z,k}(x)&\approx\frac{{\cal A}_k}{k}{\cal E}_{0x}(e^{i{\cal A}_k x}-e^{-i{\cal A}_k x})\approx i2\frac{{\cal A}_k^2}{k}{\cal E}_{0x}x,\nonumber\\
    \tilde{H}_{y,k}(x)&\approx\frac{{\omega \epsilon_{\rm I} }}{k}{\cal E}_{0x}(e^{i{\cal A}_k x}+e^{-i{\cal A}_k x})\approx2\frac{{\omega \epsilon_{\rm I} }}{k}{\cal E}_{0x}.
\end{align} 
Thereby, $H_y\equiv H_{0y}$ and $E_x\equiv E_{0x}$ are uniform across the thin insulator layer.
On the other hand, when $x > d_F$ and $x < -d_F$, the electromagnetic fields
\begin{align}
    \tilde{E}_{x,k}(x)&=\frac{\epsilon_{\rm I}}{\epsilon_0+i\sigma_c/\omega}\tilde{E}_{x,k}(\pm d_F^{\mp})e^{\pm i{\cal B}_k (x\mp d_F)},\nonumber\\
     \tilde{E}_{z,k}(x)&=\tilde{E}_{z,k}(\pm d_F)e^{\pm i{\cal B}_k (x\mp d_F)},\nonumber\\
    \tilde{H}_{y,k}(x)&=\tilde{H}_{y,k}(\pm d_F)e^{\pm i{\cal B}_k (x\mp d_F)},
    \label{H_y_in_NM}
\end{align}
where ``$+$" and ``$-$" stand for $x>d_F$ and $x<-d_F$, respectively. Therefore, $H_y\approx H_{0y}e^{-{\rm sgn}(x)(x-{\rm sgn}(x) d_F)/\delta}$ outside of the insulator.

\begin{figure}[htp]
    \centering
    \includegraphics[width=1\linewidth]{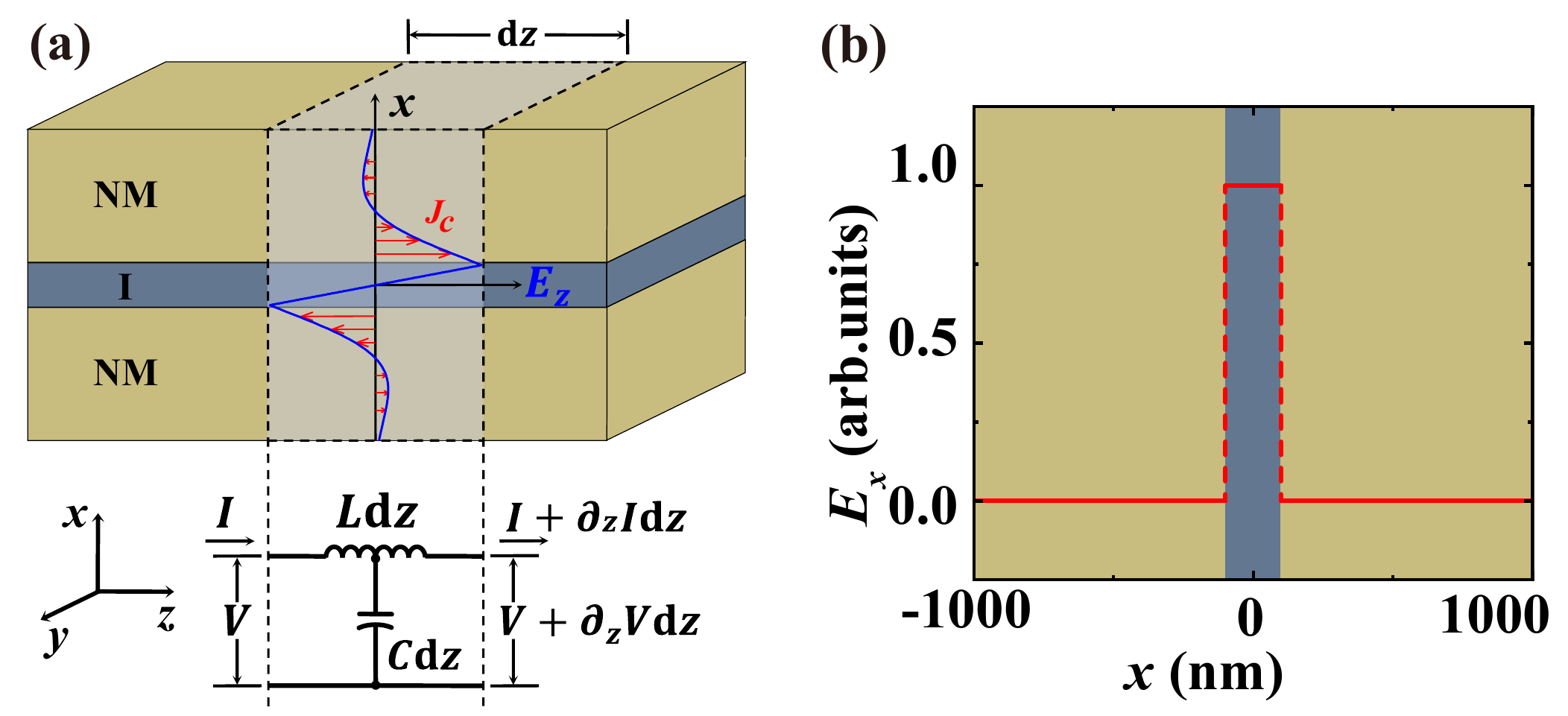}
    \caption{(a) Conversion from the coupled Maxwell's and Landau-Lifshitz equations to the transmission line model. In the NM$|$I$|$NM heterostructure, the electric field $E_z$ shows the skin effect. (b) Discontinuous distribution of the electric field $E_x$ across the thickness of the NM$|$I$|$NM heterostructure.}
    \label{figure2}
\end{figure}

The electric field can be either in-plane $E_z\hat{\bf z}$ or out-of-plane $E_x\hat{\bf x}$. The in-plane electric field $E_z\hat{\bf z}$ is continuous at the interfaces, penetrating and driving the eddy current in the NMs. Figure~\ref{figure2}(a) addresses the skin effect of the electric field $E_z$, generated by waves propagating along the $\hat{\bf z}$-direction, with a finite leakage into the adjacent NMs. Such electric field drives the eddy currents that exhibit an oscillatory decay over the skin depth $|\delta|$ from the NM$|$I interfaces. On the other hand, the out-of-plane electric field $E_x\hat{\bf x}$ induces charge accumulation at the upper and lower NM layers, which generates the voltage across the insulator. Due to the charge accumulation, this electric field vanishes inside the NMs, which only exist in the insulator layer. As depicted in Fig.~\ref{figure2}(b), the $x$-component of the electric field is discontinuous because of the discontinuity of the dielectric properties across the interfaces, which results in the accumulation of bound charges.

With the photon dispersion \eqref{photon_dispersion}, we now quantize the photon magnetic fields in terms of operators.  To this end, we interpret our model by LC circuits, as shown in Fig.~\ref{figure2}(a). We note again that since the NM$|$I$|$NM heterostructure is isotropic in the $y$-$z$ plane, without losing generality, we focus on the wave propagation along the $\hat{\bf z}$-direction. To convert our model system to the equivalent LC circuit representation, we introduce the voltage $V$ across the thickness of the heterostructure and total current $I$ per unit length along the  $\hat{\bf y}$-direction. 
From the spatial profile of the electric fields in Fig.~\ref{figure2}(a) and (b), the voltage is only contributed by the out-of-plane $x$-component of the electric field, i.e., $V=\int{\rm d}x E_x=2E_{0x}d_F$. The in-plane component of the electric field drives the eddy current. The total current $I=\oint {\bf B}\cdot d{\bf l}$ follows Amp\`ere's circuital Law, for which we select an integration route traversing the insulator in the $x$-$y$ plane. The width of this route along the $\hat{\bf y}$-direction is set to the unit length and extends to infinite along the $\hat{\bf x}$-direction. Considering the localization of the magnetic stray field, we arrive at $I=H_{0y}$. 

The dynamics of the voltage and current are not independent but interplay with each other by the electromagnetic dynamics.
An integration over Maxwell's equation $\nabla \times \mathbf{E} = -\partial\mathbf{B}/\partial t$ leads to 
\begin{equation}
    \int^{+\infty}_{-\infty}{\rm d}x\partial_z E_x=-\mu_0\int^{+\infty}_{-\infty}{\rm d} x \dot{H}_y\label{L0},
\end{equation}
where from Eq.~\eqref{H_y_in_NM} $\int_{-\infty}^{-d_F} {\rm d} x H_y=\int_{d_F}^{+\infty} {\rm d}x H_{y}=H_{0y}\delta$. Then Eq.~\eqref{L0} is simplified to be 
\begin{align}
    2\partial_z E_{0x}d_F=-2\mu_0\dot{H}_{0y} d_F(1+\delta/d_F).
    \label{quantization_1}
\end{align}
Substituting $E_{0x}$ and $H_{0y}$ with $V$ and $I$, Eq.~\eqref{quantization_1} becomes 
\begin{equation}
    \partial_z V=-2\mu_0d_F(1+{\delta}/{d_F})\dot{I}.
    \label{L1}
\end{equation}
Comparing Eq.~\eqref{L1} with the circuit equation $\partial_z V = -L \dot{I}$, we find the effective inductance 
\begin{align}
    L=2\mu_0d_F(1+\delta/d_F),
    \label{inductance}
\end{align}
which is not purely real since $\delta$ is a complex number. Additionally, substituting $E_{0x}$ and $H_{0y}$ with $V$ and $I$ to Maxwell's equation $\partial_z H_{0y}=-\epsilon_{\rm I} \dot{E}_{0x}$ yields $\partial_z I=-\epsilon_{\rm I}/(2d_F) \dot{V}$. Comparing again with the circuit equation $\partial_z I=-C \dot{V}$, we find the effective capacitance 
\begin{align}
    C=\epsilon_{\rm I}/(2d_F).
    \label{capacitance}
\end{align}

With the effective inductance \eqref{inductance} and capacitance \eqref{capacitance}, we are now allowed to express the Hamiltonian of NM$|$I$|$NM heterostructure to the effective one 
\begin{equation}
    \hat{H}=\int {\rm d}z~\frac{1}{2}( L\hat{I}^2+C \hat{V}^2).
\end{equation}
In the wave-vector space, ${\bf k}=k\hat{\bf z}$, and 
\begin{equation}
    \hat{H}=\int {\rm d}k~\frac{1}{2}\left(L \hat{I}_k \hat{I}_{-k}+\frac{CL^2}{k^2}\dot{\hat{I}}_{k}\dot{\hat{I}}_{-k}\right),
    \label{quantization_2}
\end{equation}
where $\hat{I}_k(t)=\int {\rm d}z/\sqrt{2\pi} \hat{I}(z,t) e^{-ikz}$.

We then define the canonical conjugate variables:  $\hat{q}_k = \hat{I}_k$ represents the generalized coordinate and $\hat{\pi}_k = CL^2\dot{\hat{I}}_k/k^2 $ represents the canonical momenta. With them, we express the effective Hamiltonian \eqref{quantization_2} of the system as
\begin{equation}
    \hat{H}=\int {\rm d}k\left(\frac{m_k}{2}\Omega_n^2 \hat{q}_k \hat{q}_{-k}+\frac{1}{2m_k}\hat{\pi}_k \hat{\pi}_{-k}\right),
\end{equation}
where $m_k=CL^2/k^2$.

We now define the annihilation and creation operators of photon modes as 
\begin{align}
    \hat{p}_{k}=\frac{1}{\sqrt{2m_k\hbar\Omega_n}}(m_k\Omega_n \hat{q}_k + i \hat{\pi}_{-k}),\nonumber\\
    \hat{p}_{k}^\dagger=\frac{1}{\sqrt{2m_k\hbar\Omega_n}}(m_k\Omega_n \hat{q}_{-k} + i \hat{\pi}_{k}),
\end{align}
which obey the bosonic commutation relation $[\hat{p}_k, \hat{p}_{k'}^\dagger] = \delta(k-k')$. Inversely, 
\begin{align}
    &\hat{q}_k=\sqrt{\frac{\hbar}{2m_k\Omega_n}}(\hat{p}_k+\hat{p}_{-k}^\dagger),\nonumber\\
    &\hat{\pi}_k=i\sqrt{\frac{m_k\hbar\Omega_n}{2}}\left(\hat{p}_{k}^\dagger-\hat{p}_{-k}\right).\nonumber
\end{align}
We note again in this quantization scheme that $\hat{q}_k$ and $\hat{\pi}_k$ are no longer Hermitian since the photon dispersion $\Omega_n$ is not purely real due to the Ohmic dissipation. 
This allows us to express the quantized magnetic field $\hat{\bf H}$ regarding the photon creation and annihilation operators, i.e.,
\begin{align}
    \hat{\bf H}&=\hat{H}_y\hat{\bf y} =\int \frac{{\rm d}k }{\sqrt{2\pi}} \hat{q}_k e^{ikz}\hat{\bf y}\nonumber\\
    &=\int \frac{{\rm d}k}{\sqrt{2\pi}}\sqrt{\frac{\hbar}{2m_k\Omega_n}}\left(\hat{p}_ke^{ikz}+\hat{p}_{k}^\dagger e^{-ikz}\right)\hat{\bf y}
    \label{H_y_operator}.
\end{align}
For the waves propagating with arbitrary wave vector ${\bf k}$, we express their magnetic field operators by extending Eq.~\eqref{H_y_operator} as
\begin{align}
     \hat{H}_y&=\int \frac{d{\bf k}}{2\pi}~ \sqrt{\frac{\hbar}{2m_k\Omega_n}}\cos\theta_{\bf k}\left( e^{i{\bf k}\cdot{\pmb \rho}}\hat{p}_{\bf k}+e^{-i{\bf k}\cdot{\pmb \rho}}\hat{p}_{\bf k}^\dagger\right),\nonumber\\
     \hat{H}_z&=-\int \frac{d{\bf k}}{2\pi}~ \sqrt{\frac{\hbar}{2m_k\Omega_n}}\sin\theta_{\bf k} \left(e^{i{\bf k}\cdot{\pmb \rho}}\hat{p}_{\bf k}+e^{-i{\bf k}\cdot{\pmb \rho}}\hat{p}_{\bf k}^\dagger\right).
     \label{magnetic_field_NM}
\end{align}

In terms of the magnetization operator $\hat{\bf M}$ \eqref{magnetization_NM} and the magnetic-field operator $\hat{\bf H}$ \eqref{magnetic_field_NM}, 
the Zeeman interaction describes the interaction between the magnon and photon 
\begin{align}
        \hat{H}_{\rm int}&=-\mu_0\int d{\bf r}\hat{\bf M}\cdot\hat{\bf H}=-\mu_0\int{\rm d}{\bf r}\hat{M}_y\hat{H}_y\nonumber\\
        &=\int d{\bf k}~g_{\bf k}\left(\hat{m}_{\bf k}\hat{p}_{\bf k}^{\dagger}+\hat{m}_{\bf k}^{\dagger}\hat{p}_{\bf k}-\hat{m}_{\bf k}^{\dagger}\hat{p}_{\bf -k}^{\dagger}-\hat{m}_{\bf k}\hat{p}_{\bf -k}\right),\nonumber
\end{align}
where the coupling constant
\begin{equation}
    g_{\bf k}/\hbar=\cos\theta_{\bf k} \frac{\hbar\gamma}{2} \sqrt{\frac{\mu_0(H_0+M_s)M_s}{2m_k\Omega_n\omega_Nd_F}}
    \label{coupling_NM}
\end{equation}
strongly depends on the propagation direction of the collective modes. Crucially, the coupling $g_{\bf k}\ne g_{\bf k}^*$ is dissipative~\cite{wang2020dissipative,level_attraction,Bimu} in that $\Omega_n$ and $\omega_N$ are not purely real.
The total Hamiltonian of the system
\begin{align}
    \hat{H}_{\rm tot}&=\int d{\bf k} \left(\hbar\omega_N\left(\hat{m}_{\bf k}^{\dagger}\hat{m}_{\bf k}+\frac{1}{2}\right)+\hbar\Omega_n\left(\hat{p}_{\bf k}^{\dagger}\hat{p}_{\bf k}+\frac{1}{2}\right)\right.\nonumber\\
&\left.+g_{\bf k}\left(\hat{m}_{\bf k}\hat{p}_{\bf k}^{\dagger}+\hat{m}_{\bf k}^{\dagger}\hat{p}_{\bf k}-\hat{m}_{\bf k}^{\dagger}\hat{p}_{\bf -k}^{\dagger}-\hat{m}_{\bf k}\hat{p}_{\bf -k}\right)\right)
\label{Hamiltonian_NM}
\end{align}
is non-Hermitian since $g_{\bf k}\ne g_{\bf k}^*$. This Hamiltonian goes beyond the perturbation theory without invoking the rotating-wave approximation. 
The $4\times 4$ matrix Hamiltonian \eqref{Hamiltonian_NM} under the ultrastrong coupling regime~\cite{Nori_ultrastrong} leads to the (formal) dispersion relation
\begin{equation}
\omega_{\pm}^2=\frac{1}{2}\left(\Omega_n^2+\omega_N^2\pm\sqrt{\left(\Omega_n^2-\omega_N^2\right)^2+16\Omega_n\omega_N\frac{g_{\bf k}^2}{\hbar^2}}\right).
\label{dispersion_NM}
\end{equation}  
By employing the Bogoliubov transformation, we can recover the dispersion relation for our system 
\begin{equation}
    \omega^4-\omega^2\left(\omega_N^2+\Omega_n^2\right)+\omega_N^2\Omega_n^2-4\omega_N\Omega_n{g_{\bf k}^2}/{\hbar^2}=0,\\
    \label{Qum_dispersion_NM}
\end{equation}
which is exactly the same as Eq.~\eqref{nodal_magnon_polariton} by the classical approach.

\subsection{Nodal magnon-photon polariton}

We now turn to address the properties of the collective modes in the NM$|$FI$|$NM heterostructure by performing the numerical calculation. We consider the typical magnetic insulator---yttrium iron garnet (YIG)---with the saturation magnetization $\mu_0 M_s=0.17$~T, intrinsic Gilbert damping coefficient $\alpha_G=10^{-4}$, and relative dielectric constant $\epsilon_{\rm FI}=8\epsilon_0$~\cite{Imaging_spin-wave_Bertelli,YIG_2}, sandwiched by two thick copper films with the conductivity $\sigma_{c}=5.96\times 10^{-7}$~$\Omega^{-1} \cdot {\rm m}^{-1}$. The bias magnetic field $\mu_0 H_0=50$~mT is applied along the in-plane $\hat{\bf z}$-direction, as shown in Fig.~\ref{model}. Ferromagnetic insulator EuS is also the material of choice that has been exploited in FI$|$SC$|$FI heterostructure~\cite{Robinson}. In the following, we address persistent non-Hermitian topological phenomena that appear in the wave vector space for the coupled magnon and photon modes, driven by the eddy current in the NMs that induces anisotropic dissipation $\alpha({\bf k})$.

\begin{figure}[htp]
\hspace{-0.5cm}\includegraphics[width=90mm]{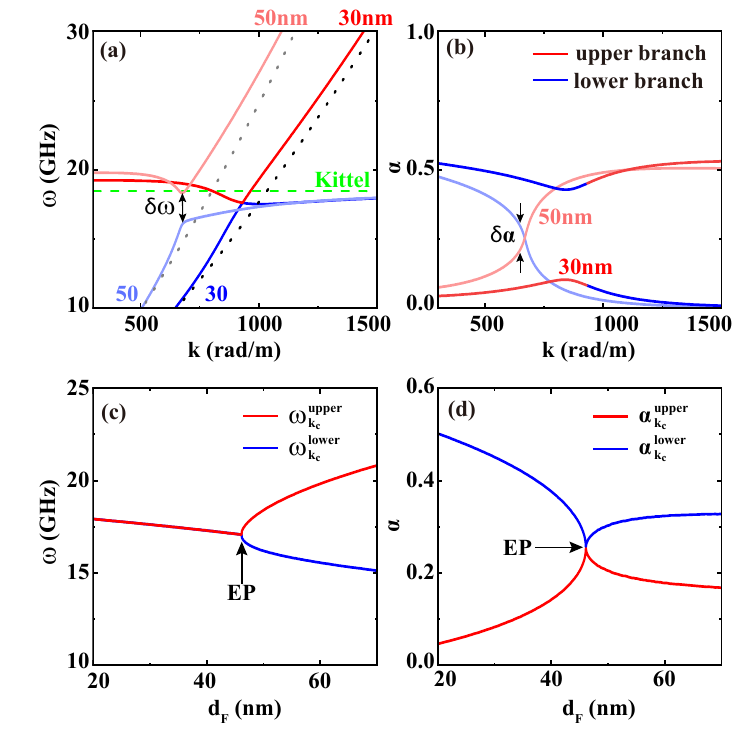}
\caption{Dispersion $\tilde{\omega}_{\bf k}=\omega_{\bf k}-i\Gamma_{\bf k}$ of collective modes in the NM$|$FI$|$NM heterostructure with exceptional nodal points. (a) plots the real part $\omega_{\bf k}$ by the red and blue solid curves when the thickness of the FI layer is $30$ and $50$~nm, respectively. The dotted curve represents the bare photon-mode frequency in the NM$|$I$|$NM heterostructure, and the dashed curve is the bare frequency of the Kittel mode. In (b), total damping for the configurations in (a) is plotted. (c) and (d) spot the EPs in the dependence on FI thickness of the frequencies and damping at ${\bf k}=k_c\hat{\bf z}$, where the gap between the two branches is minimized. }
\label{fig2}
\end{figure}

For the collective modes propagating parallel to the magnetization with $\theta_{\bf k}=0$,  we numerically solve the eigenfrequency $\tilde{\omega}_{\bf k}=\omega_{\bf k}-i\Gamma_{\bf k}$ according to Eq.~\eqref{nodal_magnon_polariton}, which contain the real $\omega_{\bf k}$ and imaginary $\Gamma_{\bf k}$ components.
Such a dispersion has two branches, namely, the upper branch $\omega^{\rm upper }_{\bf k}$ and the lower branch $\omega^{\rm lower }_{\bf k}$, with the frequency gap $\delta\omega=\min(\omega^{\rm upper }_{\bf k}-\omega^{\rm lower }_{\bf k})$ between the two branches minimized at the wavevector ${\bf k}=k_c\hat{\bf z}$, as shown in Fig.~\ref{fig2}(a). The upper branch by the red curves approaches the photon-mode frequency (the dotted curves) for large $k$ and recovers to the Kittel frequency (the dashed curves) when $k\rightarrow0$. 
On the other hand, the lower branch by the blue curves reaches the Kittel frequency for large $k$ and recovers to the photon mode when $ k\rightarrow 0$.
The frequency gap $\delta \omega$ between these two branches is sensitive to the thickness $d_F$ of the FI.
When $d_F=50$~nm, these two branches are gaped with $\delta\omega>0$; nevertheless, when $d_F=30$~nm they are gapless with $\delta\omega=0$ at a finite wave vector.  
Analogous to the frequency gap $\delta\omega$, the damping gap $\delta\alpha=\alpha_{k_c}^{\rm upper}-\alpha_{k_c}^{\rm lower}$ is defined as the difference of the damping of the two branches at ${\bf k}=k_c\hat{\bf z}$, as plotted in Fig.~\ref{fig2}(b). In contrast to the real part $\omega_{\bf k}$ of the frequency, the total damping $\alpha_{\bf k}=\Gamma_{\bf k}/\omega_{\bf k}$ of the upper and lower branches are gapless with $\delta \alpha=0$ when $d_F=50$~nm but are gapped with $\delta \alpha>0$ when $d_F=30$~nm [Fig.~\ref{fig2}(b)].

Thereby, we expect a specific FI thickness $d_F$ exists that renders the simultaneous vanished $\delta\omega$ and $\delta\alpha$, implying the collapse of two collective modes with the same real and imaginary eigenvalues. 
 Such a ``critical point" implies the existence of the EPs or the exceptional nodal phase since it appears in the wave vector space. To trace the EPs, we plot in Fig.~\ref{fig2}(c) and (d) for the thickness dependence of $\omega_{k_c}$ and $\alpha_{k_c}$ of the upper and lower branches with $d_F\in [20,70]$~nm. The EP appears when $d_F\approx46$~nm, which is experimentally feasible.

It appears that the emergence of the EPs needs the fine-tuning of parameters. For particular propagation directions, this is indeed the case, but the EPs can appear at other wave numbers. Although the damping of the bare photon and magnon modes is independent of wave propagation direction $\theta_{\bf k}$, the damping of collective modes depends on $\theta_{\bf k}$ since the magnon-photon coupling $g_{\bf k}$ [Eq.~\eqref{coupling_NM}] is anisotropic. According to Eq.~\eqref{coupling_NM}, when $\theta_{\bf k}=\pi/2$, $g_{\bf k}=0$, so the photon and magnon modes decouple; the coupling is maximal when $\theta_{\bf k}=0$.
Accordingly, although it appears above that the EPs depend on the specific thickness of the FIs, the anisotropy in the eddy-current-induced damping may render the EPs to appear persistently at the other wave vectors when the collective modes propagate along other directions, as shown below.

In Fig.~\ref{fig3}(a), we compare the dispersion of the collective modes when the spin waves propagate parallel ($\theta_{\bf k}=0$)  or skew ($\theta_{\bf k}=\pi/6$) to the saturation magnetization when  $d_F=50$~nm. In this configuration, when $\theta_{\bf k}=0$, the upper and lower branches of the collective modes are close yet separated by a gap, while when  $\theta_{\bf k}=\pi/6$, the two branches intersect without any gap. Figure~{\ref{fig3}}(b) presents the associated total damping: when $\theta_{\bf k}=0$, the damping of the two collective modes are the same, while they become different when $\theta_{\bf k}=\pi/6$. These imply the existence of the EPs in the wave vector space since the wave vector continuously changes. 
Indeed, we demonstrate in Fig.~{\ref{fig3}}(c) and (d) the existence of closure of the frequency gap and damping gap simultaneously at $\theta_{\bf k}\approx\pi/10$. 

\begin{figure}[htp]
\hspace{-0.5cm}\includegraphics[width=91mm]{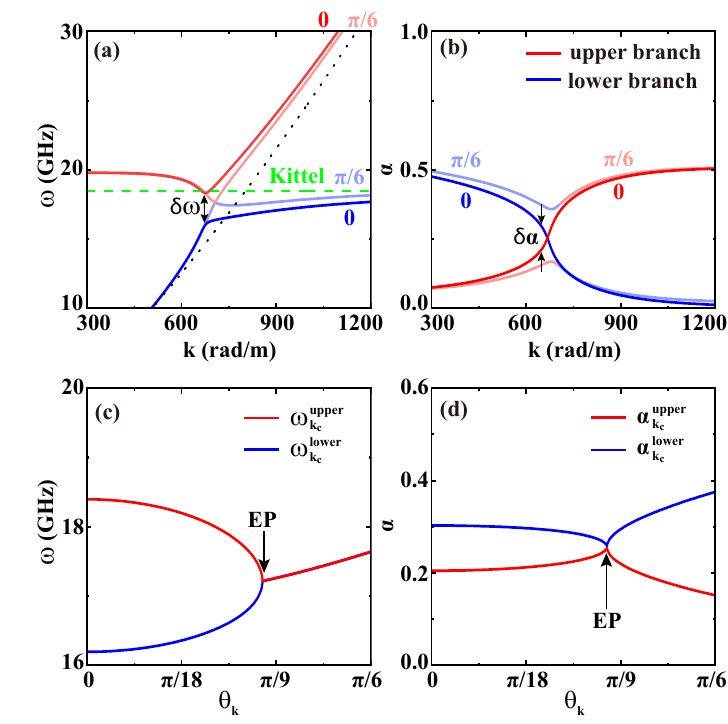}
\caption{Dispersion $\tilde{\omega}_{\bf k}=\omega_{\bf k}-i\Gamma_{\bf k}$ of the collective modes propagating in different directions $\theta_{\bf k}$ in the NM$|$FI$|$NM heterostructure with the thickness $d_F=50$~nm of the FM layer.
(a) plots the real part $\omega_{\bf k}$ when  $\theta_{\bf k}=\{0,~\pi/6\}$. The dotted curve represents the bare photon-mode frequency in the NM$|$I$|$NM heterostructure, and the green dashed curve is the bare frequency of Kittel mode. (b) plots the total damping for the configurations in (a). (c) and (d) spot the EPs in the dependence on angle ${\theta}_{\bf k}$ of the frequencies and net dampings at the wave vectors minimizing the dispersion gap. }
\label{fig3}
\end{figure}

The conditions of EPs with ultrastrong coupling go beyond the perturbation theory and differ from those with strong coupling.  
In the context of the $2\times 2$ matrix Hamiltonian after the rotating-wave approximation under the strong coupling, the condition for the emergence of EPs is given by~\cite{yu2024nonhermitian}
\begin{equation}
    (\omega_N-\Omega_n)^2+4g_{\bf k}^2=0.\label{EP_emerge}
\end{equation} 
However, when substituting the parameters in our system with which EPs are unexpected to emerge, Eq.~\eqref{EP_emerge} is inaccurate in predicting the occurrence of EPs.
Considering 
the Hamiltonian Eq.~\eqref{dispersion_NM} the EPs appear when 
\begin{align}
    \left(\Omega_n^2-\omega_N^2\right)^2+16\Omega_n\omega_Ng_{\bf k}^2/\hbar^2=0.
\end{align}
The emergence of such EPs is a joint effect of magnon ${\rm Im}\omega_N({\bf k})$ and photon ${\rm Im}\Omega_n({\bf k})$ dissipations and dissipative coupling $g_{\bf k}\ne g^*_{\bf k}$. This is very different from the EPs that appear in the ${\cal PT}$-symmetry model, where with balanced gain and loss rate $\kappa$ and coherent coupling $h_{\bf k}=h^*_{\bf k}$, the EPs appear when $\hbar \kappa=|h_{\bf k}|$. Although the mechanism is different, inserting the gain to the photon or magnon changes the properties of $\omega_N$, $\Omega_N$, and $g_{\bf k}\propto 1/\sqrt{\omega_N\Omega_n}$ simultaneously, which can thereby tune the realization condition and properties of the EPs.

\begin{figure}[htp!]
\hspace{-0.5cm}\includegraphics[width=90mm]{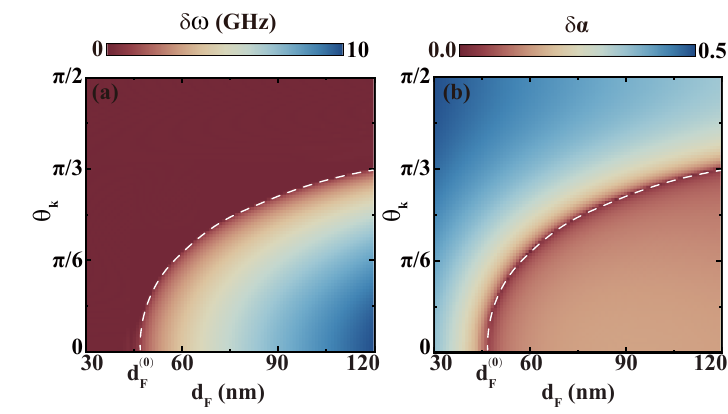}
\caption{Dependence on the FI thickness $d_F$ and propagation direction $\theta_{\bf k}$ of $\delta\omega=\min(\omega^{\rm upper }_{\bf k}-\omega^{\rm lower }_{\bf k})$ [(a)] and $\delta\alpha=\alpha_{{\bf k}_c}^{\rm upper}-\alpha_{{\bf k}_c}^{\rm lower}$ [(b)]. The dashed curves label the position of the EPs in the parameter space.}
\label{fig4}
\end{figure}

Figure~\ref{fig4} addresses the persistent existence of the EPs in the wave-vector space when the thickness of the FIs exceeds a critical value $d_F^{(0)}$, as shown in Fig.~\ref{fig4}(a) and (b). The dashed curve indicates the EPs with both gapless frequency and damping of the collective modes. In other words,  by tuning the parameters $\{d_F,H_0,\theta_{\bf k}\}$ of our system, we can discover exceptional lines, as seen in Fig.~\ref{fig4}. Furthermore, we are allowed to tune the direction of the external bias magnetic field $H_0$ in our system to persistently discover the position of EPs at a particular propagation direction. Closure and reopening of the ``frequency gap" $\delta\omega$ and ``damping gap" $\delta\alpha$ happen across the exceptional points/lines, which may be treated as a ``topological" phase transition since one may define the energy vorticity as a topological characterization for such collective modes. In the phase space spanned by $\{\delta\omega,\delta\alpha\}$, the EPs are singularities around which the energy vorticity does not vanish~\cite{yu2024nonhermitian}.

EPs of magnons have been observed in ferromagnetic multilayers~\cite{liu2019observation} as well as in systems combining magnon with cavity photons through the dipolar interaction~\cite{zhang2019experimental,zhang2017observation,harder2017topological,zhang2019higher}.
The emergence of EPs in the passive magnonic $\mathcal{PT}$-symmetric device reported in FM$|$NM$|$FM heterostructure~\cite{ liu2019observation} is realized through fine-tuning the gain, dissipation, and the RKKY interaction by changing the NM thickness.

The coalescence of the two modes is different from the level attraction phenomena~\cite{wang2020dissipative,level_attraction,Bimu}. The level attraction between two modes was observed in dissipative coupling systems~\cite{wang2020dissipative,Bimu,level_attraction} when the absolute value of the dissipative coupling strength is larger than the difference of their eigenfrequencies. Within the realm of level attraction, the eigenfrequencies of the two modes become degenerate while their dissipation rates differ. In contrast, the coalescence of two modes is the hallmark of EPs, with both the eigenfrequencies and dissipation rates of the two modes being the same. In this sense, the EPs could be used for the synchronization of two oscillators.

\section{SC$|$FI$|$SC Josephson junction}
\label{Sec_IV}

\subsection{Classical approach}

As an extension of the above formalism, we consider the SC$|$FI$|$SC heterostructure. We find most formalism with NMs addressed above can recover that with SCs when replacing the skin depth $\delta$ by London's penetration depth $\lambda$, although the former is non-Hermitian and the latter is Hermitian. We assume the temperature $T\lesssim 0.85 T_c$, at which the conductivity is mainly contributed by the superfluid, such that we can safely disregard the nondesirable thermal effect that contributes to the small additional damping of magnons.

Since $1/\lambda^2\gg k_0^2$ when $\omega\lesssim 100$~GHz, ${\cal B}_k\approx i/\lambda $.  On the other hand, $|{\cal A}_k|\sim k$ such that for a thin FI,  $|{\cal A}_kd_F|\sim kd_F\ll 1$. Accordingly, we expand $e^{\pm i{\cal A}_kd_F}=1\pm i{\cal A}_kd_F$, which simplifies  Eq.~\eqref{amplitudes_magnetic_field}  to be 
    \begin{align}
    &{\cal H}_{x}= -\dfrac{1-k_0^2d_F\lambda}{1-{\cal A}_k^2d_F\lambda}{\cal M}_{x}\approx-{\cal M}_{x},\nonumber\\
   & {\cal H}_{y}= -\dfrac{d_F}{k_0^2\lambda+{\cal A}_k^2d_F}\left(k_0^2-\frac{d_F}{\lambda+d_F}k_y^2\right){\cal M}_{y}.\label{SC}
\end{align}
This implies that the out-of-plane $B_x$ is nearly zero in the FI. This is reasonable because the magnetic induction cannot exist inside the SC, and $B_x$ is continuous across the SC$|$FI interfaces. The SC then mainly modulates the in-plane magnetic field along the $\hat{\bf y}$-direction.
On the other hand, the electric field inside the FI according to Eqs.~\eqref{Ex}, \eqref{Ey}, and \eqref{Ez} is reduced to 
    \begin{align}
    {\tilde E}_x&=-\frac{\omega\mu_0k_zd_F}{{\cal A}_k^2d_F+k_0^2 \lambda}{\cal M}_{y},\nonumber\\
    {\tilde E}_y&=-\omega\mu_0\lambda d_F k_z {\cal M}_{x}-i\frac{\lambda}{d_F+\lambda}\frac{\omega \mu_0 d_F}{{\cal A}_k^2d_F+k_0^2\lambda}k_yk_z{\cal M}_{y}x,\nonumber\\
    {\tilde E}_z&=\omega\mu_0\lambda d_F k_y {\cal M}_{x}-i\omega\mu_0\dfrac{(k_0^2-\frac{d_F}{\lambda+d_F}k_y^2)\lambda}{k_0^2\lambda+{\cal A}_k^2d_F}{\cal M}_{y}x.
    \label{Edis}
\end{align}

Combining with the linearized Landau-Lifshitz-Gilbert equation~\eqref{LLG},  the dispersion relation of the collective mode is governed by 
\begin{equation}
    k^2+K_{\omega}\frac{d_F}{d_F+\lambda}k_y^2=k_0^2\left(1+K_\omega+\frac{\lambda}{d_F}\right).
    \label{dpSC}
\end{equation}
We note that replacing $\lambda$ in Eq.~\eqref{dpSC} with $\delta$ reduces exactly this characteristic equation of the SC$|$FI$|$SC Josephson junction to that of the NM$|$FI$|$NM heterostructure \eqref{nodal_magnon_polariton}. 
From Eq.~\eqref{dpSC}, we obtain two real solutions
\begin{equation}
    \omega^2_{u (l)}({\bf k})=\frac{1}{2}\left(\Omega_s^2(k)+\omega_S^2\pm \sqrt{\left(\Omega_s^2(k)-\omega_S^2\right)^2+4\Delta_{\bf k}^2}\right),
\label{SC_dispersion}
\end{equation}
where 
\begin{align}
\Omega_s(k)&=\sqrt{{d_F}/({d_F+\lambda})}ck,\nonumber\\
\omega_S&=\mu_0\gamma\sqrt{(H_0+M_s)(H_0+M_sd_F/(d_F+\lambda))},
\label{Swihart_frequency}
\end{align}
are, respectively, the frequency of the Swihart mode~\cite{swihart} between two SCs (refer to Appendix~\ref{appendix_Swihart}) and the ``renormalized" FMR frequency by the SCs~\cite{zhou2023gating,Silaev}, and 
\begin{align}
    \Delta^2_{\bf k}=\cos^2\theta_{\bf k}\Omega_s^2(k)\mu_0^2\gamma^2M_s(H_0+M_s)\frac{d_F}{d_F+\lambda}
\end{align}
relates to the anisotropic coupling between the Swihart and Kittel modes. Here $c=1/{\sqrt{\mu_0\epsilon_{\rm FI}}}$ is the light velocity in the FI.

The maximal value of $\Delta_{\bf k}^2$ appears when $\cos^2\theta_{\bf k}=1$, i.e., the modes propagating parallel to ${\bf M}_s$, resulting in the largest splitting. While when the collective modes propagate perpendicular to the external magnetic field or ${\bf M}_{s}$, $\cos^2\theta_{\bf k}=0$, the Swihart mode and Kittel mode decouples with  
\begin{equation}
    \omega_u=\left\{
    \begin{aligned}
    &\omega_S,~~~~~~k_y\leq k_c\\
    &\Omega_s(k_y),~k_y>k_c
\end{aligned}\right.,~\omega_l=\left\{
    \begin{aligned}
    &\Omega_s(k_y),~k_y\leq k_c\\
    &\omega_S,~~~~~~~k_y>k_c
\end{aligned}\right.,
\end{equation}
where $k_c=\sqrt{1+\lambda/d_F}\omega_S/c$ is the crossing point of the two modes. 
These features in the coupling between 
the two modes can be understood from the Zeeman interaction $-\mu_0{M}_y{H}_y$.
When $\mathbf{k}$ is perpendicular to ${\bf M}_\mathrm{s}$, substituting $\Omega_s$ into Eq.~\eqref{SC} yields $H_y = 0$. Consequently, the Zeeman interaction between the magnon and photon modes vanishes. On the other hand, when $\mathbf{k}$ is parallel to ${\bf M}_\mathrm{s}$, $k_y = 0$, and $H_y$ reaches its maximum according to Eq.~\eqref{SC}. As a result, the Zeeman interaction between the two modes is maximized in the bulk volume configuration.

When $k\rightarrow 0$, the frequencies of the two branches are reduced to
\begin{align}
\omega_u&=\omega_S\nonumber,\\
\omega_l&=\sqrt{\frac{H_0(d_{F}+\lambda)+M_sd_F\sin^2\theta_{\bf k}}{H_0(d_F+\lambda)+M_sd_F}\frac{d_F}{d_F+\lambda}}ck\rightarrow 0.
\end{align}
Accordingly, $\omega_u(k\rightarrow 0)$ is the shifted FMR~\cite{zhou2023gating,silaev2023ultrastrong}, while $\omega_l(k\rightarrow 0)$ renders the Swihart photon modes anisotropic and reduces the group velocity. 
On the other hand, when $k$ is large, i.e., $ck\gg\mu_0\gamma(H_0+M_s)$, 
\begin{align}
\omega_{u}&=\sqrt{\frac{d_F}{d_F+\lambda}(c^2 k^2+\mu_0^2\gamma^2(H_0+M_s)M_s\cos^2\theta_{\bf k})}\nonumber\\
&\approx\sqrt{\frac{d_F}{d_F+\lambda}}ck=\Omega_s(k), \nonumber\\
\omega_{l}&=\mu_0\gamma\sqrt{(H_0+M_{s})\left(H_0+\frac{d_F}{d_F+\lambda}M_{s}\sin^2\theta_{\bf k}\right)}.
\end{align}
Thereby, $\omega_u$ is switched to the Swihart mode, while $\omega_l$ becomes the shifted FMR.

\subsection{Quantum approach}

In the SC$|$FI$|$SC Josephson junction, the photon modes are well confined between two SC layers, acting as an ultrathin microwave waveguide without introducing dissipation. Such photon modes are described by the Swihart mode~\cite{swihart}. 
In the long wavelength limit, similar to the NM case \eqref{kittel_NM}, the eigenfrequency of magnon is also given by Kittel's  formula~\cite{Kittel_mode} 
\begin{equation}
    \omega_S=\mu_0\gamma\sqrt{(H_0+N_{xx}M_{s})(H_0+N_{yy}M_{s})},
    \label{kittel}
\end{equation}
but the demagnetization factors are renormalized 
 to be $N_{xx}=1$ are  $N_{yy}=d_F/(d_F+\lambda)$ by the SCs, as shown explicitly in Appendix~\ref{demagnetization_factor}. The interaction between magnon and Swihart photon is described by the Zeeman interaction
\begin{equation}
    \hat{H}_{\rm int}=-\mu_0\int{\rm d}{\bf r}~ \hat{\bf M}\cdot \hat{\bf H}_{\rm Sw},\nonumber
\end{equation}
where $\hat{\bf M}$ is the magnetization operator for the magnon mode and $\hat{\bf H}_{\rm Sw}$ is the magnetic-field operator for the Swihart mode.

Combining the Landau-Lifshitz equation and normalization relation $\int d\mathbf{r}\left({M}_x{M}_y^{*}-{M}_x^{*}{M}_y \right)=-i/2$~\cite{FMR_Walker,Spinwaveexcitations}, we find the normalized amplitudes
\begin{align}
    &{\cal M}_{x}=i\sqrt{\frac{\omega_S}{8d_F\mu_0\gamma(H_0+M_{s})}},\nonumber\\
    &{\cal M}_{y}=-\sqrt{\frac{\mu_0\gamma(H_0+M_{s})}{8d_F\omega_S}}.\nonumber
\end{align}
The magnetization is then quantized as  
\begin{align}
    \hat{M}_x&=\int\frac{d{\bf k}}{2\pi} \left(-i\sqrt{\frac{\hbar M_s\omega_S}{4d_F\mu_0(H_0+M_s)}}e^{i{\bf k}\cdot{\pmb \rho}}\hat{m}_{\bf k}+{\rm H.c.}\right),\nonumber\\
    \hat{M}_y&=\int\frac{d{\bf k}}{2\pi} \left(\sqrt{\frac{\hbar\mu_0\gamma^2(H_0+M_s)M_s}{4d_F\omega_S}}e^{i{\bf k}\cdot{\pmb \rho}}\hat{m}_{\bf k}+{\rm H.c.}\right).
\end{align}
According to Appendix~\ref{appendix_Swihart}, the magnetic field is quantized according to  
\begin{align}
     \hat{H}_{{\rm Sw},y}&\approx \int \frac{d{\bf k}}{2\pi}~ \left(\frac{\Omega_s}{2k}\sqrt{\frac{\epsilon_{\rm FI}\hbar\Omega_s}{d_F}}\cos\theta_{\bf k} e^{i{\bf k}\cdot{\pmb \rho}}\hat{p}_{\bf k}+{\rm H.c.}\right),\nonumber\\
     \hat{H}_{{\rm Sw},z}&\approx \int \frac{d{\bf k}}{2\pi}~ \left(-\frac{\Omega_s}{2k}\sqrt{\frac{\epsilon_{\rm FI}\hbar\Omega_s}{d_F}}\sin\theta_{\bf k} e^{i{\bf k}\cdot{\pmb \rho}}\hat{p}_{\bf k}+{\rm H.c.}\right),
\end{align}
where $\hat{p}_{\bf k}$ is the annihilation operator of Swihart photons. 
These quantizations lead to  
\begin{align}
        \hat{H}_{\rm int}&=-\mu_0\int d{\bf r}\hat{M}_y\hat{H}_{{\rm Sw}, y}\nonumber\\
        &=\int d{\bf k}~g_{\bf k}\left(\hat{m}_{\bf k}\hat{p}_{\bf k}^{\dagger}+\hat{m}_{\bf k}^{\dagger}\hat{p}_{\bf k}-\hat{m}_{\bf k}^{\dagger}\hat{p}_{\bf -k}^{\dagger}-\hat{m}_{\bf k}\hat{p}_{\bf -k}\right),\nonumber
\end{align}
where the coupling constant
\begin{equation}
    g_{\bf k}/\hbar=\cos\theta_{\bf k}\frac{\mu_0\gamma\Omega_s}{2|{\bf k}|}\sqrt{\frac{(H_0+M_s)M_s\Omega_s\mu_0\epsilon_{\rm FI}}{\omega_S}}
    \label{coupling_strength_SC}
\end{equation}
strongly depends on the propagation direction of the collective modes. 
The total Hamiltonian of the system reads 
\begin{align}
    \hat{H}_{\rm tot}&=\int d{\bf k} ~\left(\hbar\omega_S\left(\hat{m}_{\bf k}^{\dagger}\hat{m}_{\bf k}+\frac{1}{2}\right)+\hbar\Omega_s(k)\left(\hat{p}_{\bf k}^{\dagger}\hat{p}_{\bf k}+\frac{1}{2}\right)\right.\nonumber\\
&\left.+g_{\bf k}\left(\hat{m}_{\bf k}\hat{p}_{\bf k}^{\dagger}+\hat{m}_{\bf k}^{\dagger}\hat{p}_{\bf k}-\hat{m}_{\bf k}^{\dagger}\hat{p}_{\bf -k}^{\dagger}-\hat{m}_{\bf k}\hat{p}_{\bf -k}\right)\right),
\label{SC_hamiltonian}
\end{align}
which goes beyond the perturbation theory.

After the Bogoliubov transformation on the Hamiltonian \eqref{SC_hamiltonian}, the eigenfrequency of the collective modes is governed by the characteristic equation 
\begin{equation}
    \omega^4({\bf k})-\omega^2\left(\Omega_s^2(k)+\omega_S^2\right)+\Omega_s^2(k)\omega_S^2-4\Omega_s(k)\omega_S{g_{\bf k}^2}/{\hbar^2}=0.
    \label{Qum_dispersion}
\end{equation}
Substitution of $\Omega_s(k)$ \eqref{Swihart_frequency} and $g_{\bf k}$ \eqref{coupling_strength_SC} into Eq.~\eqref{Qum_dispersion} yields 
\begin{align}
    &\omega^4-\omega^2\left(\frac{d_F}{d_F+\lambda}\frac{k^2}{\mu_0\epsilon_{\rm FI}}+\omega_S^2\right)+\omega_S^2\frac{d_F}{d_F+\lambda}\frac{k^2}{\mu_0\epsilon_{\rm FI}}\nonumber\\
    &-\mu_0^2\gamma^2(H_0+M_s)M_s\left(\frac{d_{F}}{d_{F}+\lambda}\right)^2\frac{k_z^2}{\mu_0\epsilon_{\rm FI}}=0,
    \label{Qum_dispersion_SC}
\end{align}
which is exactly the same as Eq.~\eqref{SC_dispersion} derived by the classical approach.

The solution of Eq.~\eqref{Qum_dispersion} is 
\begin{equation}
    \omega_{u(l)}^2=\frac{1}{2}\left(\Omega_s^2(k)+\omega_S^2\pm\sqrt{\left(\Omega_s^2(k)-\omega_S^2\right)^2+16\Omega_s(k)\omega_S\frac{g_{\bf k}^2}{\hbar^2}}\right).
\end{equation}
When $\Omega_s(k)=\omega_S$, the magnon and Swihart modes cross with each other, and the solutions at the crossing point $\omega^{*}_{u(l)}=\sqrt{\omega_S^2\pm2\omega_Sg_{\bf k}/\hbar}$ defines the anti-crossing gap as 
\begin{align}
    \Delta\omega&=\omega^{*}_{u}-\omega^{*}_{l}\nonumber\\
    &=\sqrt{\omega_S^2+2\omega_Sg_{\bf k}/\hbar}-\sqrt{\omega_S^2-2\omega_Sg_{\bf k}/\hbar}.
\end{align}
When $g_{\bf k}/\hbar\ll\omega_S$, $\Delta\omega\approx 2g_{\bf k}/\hbar$ is the same as the solution obtained by the perturbation theory~\cite{cavity_magnonics}, which, however, breaks down when $g_{\bf k}/\hbar$ is comparable to the bare magnon frequency $\omega_S$. We also note that the stable solution only exists when $2g_{\bf k}/\hbar<\omega_S$. Because $2g_{\bf k}/\hbar=\cos\theta_{\bf k}\mu_0\gamma\sqrt{(H_0+M_s)M_sd_F/(d_F+\lambda)}<\mu_0\gamma\sqrt{(H_0+M_s)(H_0+M_sd_F/(d_F+\lambda))}=\omega_S$, we confirm that the solution of \eqref{Qum_dispersion_SC} is always stable.

 The gap $\Delta\omega$ approaches a constant when $d_F\gg\lambda$. In this limit the frequency of the Swihart mode $\Omega_s\rightarrow c|{\bf k}|$, the ``renormalized" Kittel frequency $\omega_S\rightarrow\mu_0\gamma(H_0+M_s)$, and the coupling constant $g_{\bf k}/\hbar\rightarrow\cos\theta_{\bf k}\mu_0\gamma\sqrt{(H_0+M_s)M_s}/2$. Accordingly, \begin{align}
    \Delta\omega|_{d_F\gg \lambda}&\rightarrow\mu_0\gamma\sqrt{(H_0+M_s)}\nonumber\\
&\times\left(\sqrt{(H_0+M_s)+\sqrt{(H_0+M_s)M_s}}\right.\nonumber\\
    &\left.-\sqrt{(H_0+M_s)-\sqrt{(H_0+M_s)M_s}}\right)
    \label{limit}
\end{align}
is free of geometry parameters.

\subsection{Anisotropic ultrastrong coupling}

\begin{figure}[htp!]
    \centering
    \includegraphics[width=1.0\linewidth]{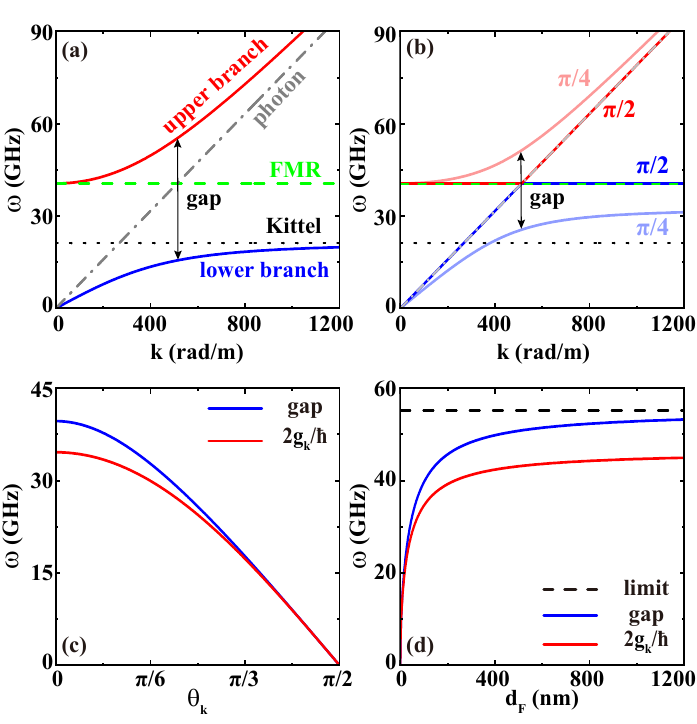}
    \caption{ Level repulsion with ultrastrong coupling between the magnon and Swihart photon modes in the SC$|$FI$|$SC Josephson junction. (a) shows the dispersion of the collective modes propagating in parallel to the saturation magnetization with $\theta_{\bf k}=0$  when the thickness $d_F=100$~nm. (b) plots the dispersion of the collective modes with different propagation directions $\theta_{\bf k}=\{\pi/4,\pi/2\}$. 
    (c) and (d) address the propagation direction and FI thickness dependencies of the gap of the level repulsion.
    }
    \label{SC1}
\end{figure}

We then address the parameter dependence in the frequencies of the two collective-mode branches $\omega_u({\bf k})$ and $\omega_l({\bf k})$ [Eq.~(\ref{SC_dispersion})] in the SC$|$FI$|$SC heterostructure,  which behave as an anti-crossing in the wave-vector space. Figure~\ref{SC1} plots the anisotropic ultrastrong coupling between the magnon mode and Swihart photon mode at different propagation directions and
addresses the wave-vector and thickness dependencies of the coupling strength to find optimal parameters. In the calculation, the thickness of YIG film $d_F=100$~nm, the saturation magnetization at low temperatures is enhanced to be $\mu_0 M_s=0.24$~T~\cite{Borst,YIG_m0}, biased by the external magnetic field $\mu_0 H_0=50$~mT, the intrinsic Gilbert damping $\alpha_G=10^{-4}$, and $\epsilon_{\rm FI}=8\epsilon_0$~\cite{YIG_2}. We use the superconducting NbN film with London's penetration depth $\lambda=80$~nm at $T=0.1T_c\sim 1$~K~\cite{NbN,NbN3,NbN2}.

When the collective mode propagates parallel to the magnetization with $\theta_{\bf k}=0$, according to Eq.~(\ref{SC_dispersion}) in the classical approach, the two branches show level repulsion with a frequency gap $\Delta\omega$, as shown in Fig.~\ref{SC1}(a). For large $k$, the upper branch by the red curve approaches the high-frequency Swihart photon mode, and when $k\rightarrow 0$, it recovers to the renormalized FMR, which is shifted giantly in comparison to the bare Kittel frequency~\cite{Silaev,zhou2023gating}.
On the other hand, the lower branch by the blue curve reaches the bare Kittel frequency for large $k$ and recovers to the Swihart photon mode when $k\rightarrow0$. The frequency gap $\Delta\omega=\omega_{u}-\omega_{l}$ is defined at the crossing point ${\bf k}= {k}_c\hat{\bf z}$ of the renormalized FMR and Swihart-mode frequency. In Fig.~\ref{SC1}(a), the gap $\Delta\omega\sim 39.7$~GHz is even larger than the bare Kittel frequency but comparable to the renormalized Kittel frequency, showing the ultrastrong coupling between the magnon and Swihart-photon modes. Furthermore, we find even large Gilbert damping $\alpha_G=0.1$ influences little the anticrossing in the ultrastrong coupling regime.

These features in the bulk-volume configuration were also revealed recently by Silaev in Ref.~\cite{silaev2023ultrastrong}. References~\cite{golovchanskiy2021ultrastrong,golovchanskiy2021approaching} found the level repulsion in the ${\rm S|F|S'|I|S''}$ multilayers, which may be interpreted as the Kittel mode with shifted FMR frequency in ${\rm S|F|S'}$ couples with the Swihart photon mode in ${\rm S'|I|S''}$, mediated by the supercurrent in ${\rm S'}$.

Reference~\cite{silaev2023ultrastrong} focuses on the specific bulk-volume configuration with $\theta_{\bf k}=0$. Here, we go beyond this configuration by considering the arbitrary propagation direction of the collective modes. As shown in Fig.~\ref{SC1}(b) with $\theta_{\bf k}\neq0$, the anti-crossing gap, or the level repulsion strongly depends on the propagation direction. When $\theta_{\bf k}=\pi/4$, the frequency gap $\Delta\omega$ at $|{\bf k}|=k_c$ becomes smaller than that in Fig.~\ref{SC1}(a). Furthermore, when the collective modes propagate normally to the magnetization with $\theta_{\bf k}=\pi/2$, the crossing of the upper and lower branches is not gapped, i.e., the magnon mode and Swihart photon mode decouples since $g_{\bf k}=0$ [Eq.~\eqref{coupling_strength_SC}]. Figure~\ref{SC1}(c) summarizes the dependence of the anti-crossing gap $\Delta\omega$ on the propagation direction $\theta_{\bf k}$. Since  the coupling constant  $2g_{k_c}/\hbar\sim \cos\theta_{\bf k}$ according to Eq.~(\ref{coupling_strength_SC}), the gap reaches its maximum when $\theta_{\bf k}=0$ and becomes zero when $\theta_{\bf k}=\pi/2$ as plotted in Fig.~\ref{SC1}(c). Furthermore, the gap $\Delta \omega>2g_{k_c}/\hbar$ when $\theta_{\bf k}\rightarrow 0$, exhibiting the feature of ultrastrong coupling, while $\Delta \omega\approx 2g_{k_c}/\hbar$ when $\theta_{\bf k}\rightarrow \pi/2$ such that the perturbation theory applies.

The thickness of the ferromagnetic film  $d_F$ strongly affects the coupling strength as well when $d_F<\lambda$, but becomes saturated to that in Eq.~\eqref{limit} when $d_F\gg \lambda$, as shown in Fig.~\ref{SC1}(d) when $\theta_{\bf k}=0$. Here, it is natural to expect the coupling strength to vanish when $d_F\rightarrow 0$ since then the FI is absent. For the thick YIG film, $d_F\gg \lambda$ increases the anti-crossing gap up to $\sim 55$~GHz, close to reaching the limit [about 56~GHz according to Eq.~\eqref{limit}].  This suggests the optimal thickness of the FI is $d_F\sim \lambda$, which depends on the temperature since $\lambda=\lambda(T=0)(1-(T/T_c)^4)$.  $\Delta \omega>2g_{k_c}/\hbar$ suggests again the perturbation theory breaks down under the ultrastrong coupling.

By changing the direction of the magnetic field while fixing the propagation direction of the collective modes, we can either maximize the coupling strength to enter the ultrastrong coupling regime or minimize it to decouple from photons, which is advantageous for broadband microwave filters by the forbidden band~\cite{dispersion_engeneering} and exploring quantum effects with high information processing efficiency and tunability~\cite{silaev2023ultrastrong,qm_effect1, qm_effect2}.

The formulation of the magnon-photon polariton in the NM(SC)$|$FI$|$NM(SC) heterostructure considers the linear regime of magnetization dynamics, similar to that of exciton-photon polariton~\cite{Hopfield}. We envision nonlinearity can appear when the number of magnon-photon polariton increases by strong microwave pumping due to the interaction between magnons~\cite{nonlinearity1,nonlinearity2}. The heterostructure might be a platform to study its condensation.

\section{Discussion and conclusion}
\label{Sec_V} 

In conclusion, we develop a non-perturbation theory using both classical and quantum approaches to investigate the collective magnon-photon modes in the ferromagnetic heterostructures composed of ferromagnetic insulators sandwiched by either normal metals or superconductors, thereby representing either dissipative or non-dissipative states of electrons. We develop a non-Hermitian quantization scheme with non-Hermitian magnetization and magnetic/electric-field operators to account for the Ohmic dissipation, due to which the coupling between magnon and photon is dissipative. A simple conversion relation between the two electronic states is revealed by finding most formalisms with normal metals can be extended to those with superconductors by simply replacing the skin depth $\delta$ with London's penetration depth $\lambda$, although the former is non-Hermitian and the latter is Hermitian. Our formalism is  ``phenomenological" in the sense that the thermal effect and interactions between magnons and other quasiparticles are accounted for by the phenomenological saturation magnetization, Gilbert damping, penetration depth, and electric conductivity that may depend on the thermal fluctuation and various interactions in different materials, which we take typical values from the experiments.

In the NM$|$FI$|$NM heterostructure, we predict the persistent realization of non-Hermitian nodal magnon-photon polariton of long wavelength, where the coupling strength is comparable to the bare magnon frequency. 
Remarkably, the exceptional points at which the eigenvalues and eigenvectors are coalescent persist over a wide range of wave vectors when the ferromagnetic layer is sufficiently thick $\sim 100$~nm, without the need for fine-tuning of parameters. The dissipationless nature of the superconductors eliminates the eddy-current-induced damping present in the normal metal case. In the SC$|$FI$|$SC Josephson junction, we find that magnon-photon coupling exhibits strong anisotropy. The coupling is ultrastrong when the electromagnetic field propagates along the direction of the saturation magnetization but vanishes when they are normal to each other due to the chirality of the dipolar stray field.

Exceptional points hold the potential application in the enhancement of sensitivity in magnonic devices, as recently observed in the enhanced magnonic frequency combs~\cite{wang2024enhancement}. Our results suggest that the common ferromagnetic heterostructures can host persistent EPs and anisotropic ultrastrong magnon-photon coupling, which may enable improved sensitivity and performance of the magnonic devices~\cite{bender2007making, heiss2012physics}. The predicted persistent nodal polariton and strongly anisotropic coupling provide new opportunities to explore the synergies between magnonics and superconducting electronics~\cite{zhang2017observation, harder2017topological, zhang2019higher, zhang2019experimental, liu2019observation, wang2023floquet, dispersion_engeneering, qm_effect2, qm_effect1}.

\begin{acknowledgments}
This work is financially supported by the National Key Research and Development Program of China under Grant No.~2023YFA1406600, the National Natural Science Foundation of China under Grants No.~12374109 and No.~52201200, as well as the startup grant of Huazhong University of Science and Technology. H.W. acknowledges support from the China Scholarship Council (CSC) under Grant No. 202206020091.
\end{acknowledgments}

\begin{appendix}

\section{Renormalized demagnetization factor in the magnetic heterostructure}
\label{demagnetization_factor}

\begin{figure}[htp]
    \centering
    \includegraphics[width=0.95\linewidth]{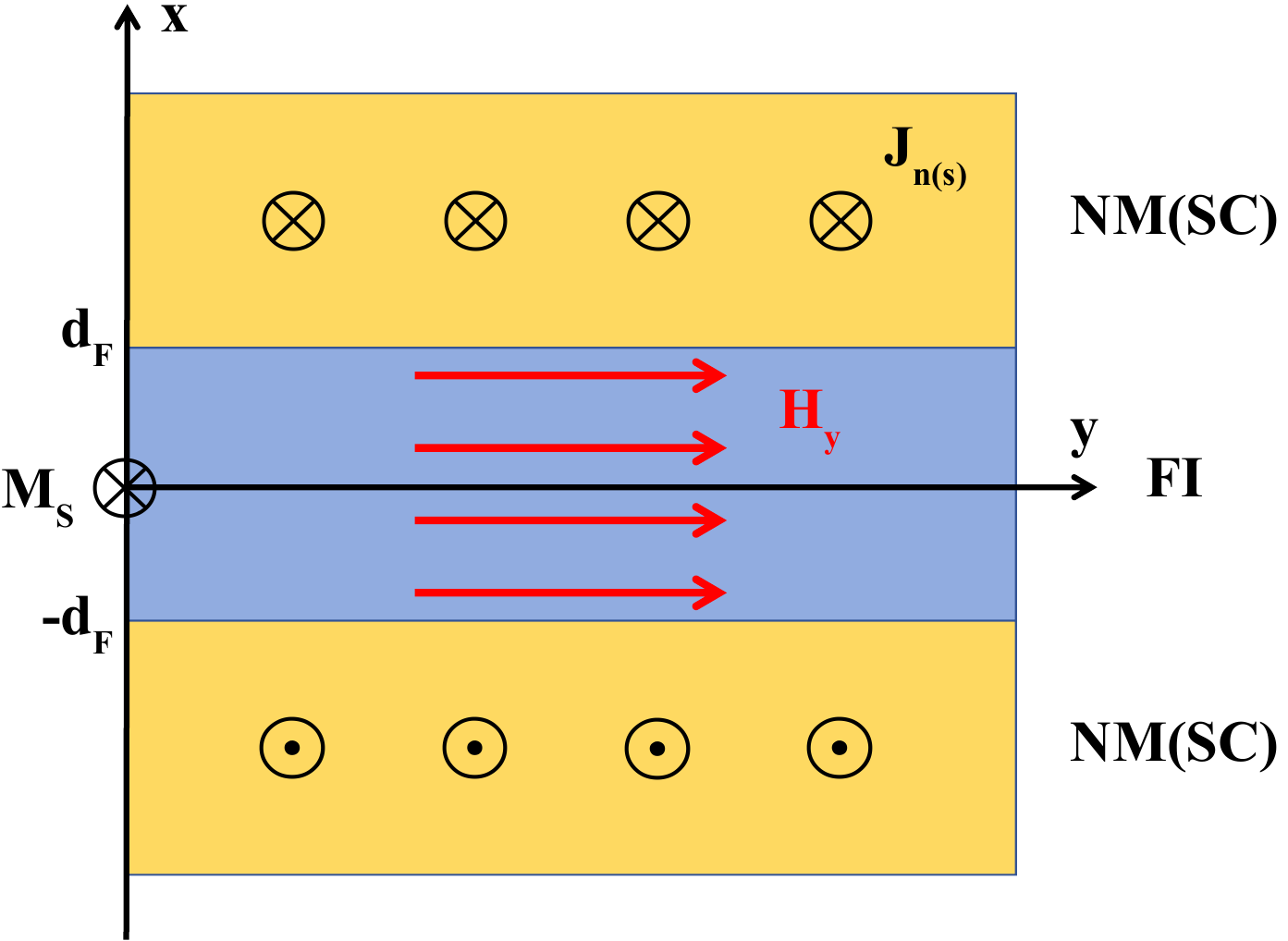}
    \caption{Configuration of the magnetic heterostructures. The current $J_{n(s)}\hat{\bf z}$ driven by the radiated electric field of ferromagnetic resonance generates the Oersted field $H_y\hat{\bf y}$ inside the ferromagnet.}
    \label{fig:3}
\end{figure}

In this appendix, we calculate explicitly the renormalized demagnetization factors in the magnetic heterostructures~\cite{zhou2023gating}.
In the configuration shown in Fig.~\ref{fig:3}, a FI of thickness $2d_F$ is sandwiched by NMs or SCs. The dynamics of ferromagnetic resonance in such NM(SC)$|$FI$|$NM(SC) heterostructure is governed by the Maxwell's equations
\begin{align}
    &\text{in FI:}~~~~~~~~~~~~\nabla^2 {\bf E}+ k_0^2 {\bf E}=-i\omega\mu_0\nabla\times{\bf M},\nonumber\\
    &\text{in NM(SC):}~~~~\nabla^2 {\bf E}+ k^2_{n(s)}{\bf E}=0,
\end{align}
where $k_0=\sqrt{\omega^2\mu_0 \epsilon_{\rm FI}}$ and $k_{n(s)}=\sqrt{\omega^2\mu_0 \epsilon_0+i\omega \mu_0 \sigma_{c(s)}}$ with $\sigma_{c(s)}$ being the conductivity of  NMs or SCs.
Combining the full solution for the electric fields $E_z(x,t)=\tilde{E}_z(x) e^{-i\omega t}$, where
\begin{align}
    &\text{in FI:}~~~~~~~~~~~~~~~~~~~~~~~~{\tilde E}_z(x)={\cal E}_{0z}e^{ik_0x}+{\cal E}_{z0}'e^{-ik_0x},\nonumber\\
    &\text{in NM(SC)}~(x>d_F)\text{:}~~~{\tilde E}_z(x)={\cal E}_{1z}e^{ik_{n(s)}x},\nonumber\\
    &\text{in NM(SC)}~(x<-d_F)\text{:}~{\tilde E}_z(x)={\cal E}_{2z}e^{-ik_{n(s)}x},\nonumber
\end{align}
with the boundary conditions, we solve all the amplitudes $\{{\cal E}_{0z},{\cal E}_{0z}',{\cal E}_{1z},{\cal E}_{2z}\}$: inside the FI,
\begin{equation}
    E_z(|x|<d_F)=-\frac{\omega\mu_0M_y\sinh (ik_0 x)}{k_0\cosh(ik_0d_F)-k_{n(s)}\sinh(ik_0d_F)};
    \label{E_z}
\end{equation}
inside the NM or SC,
\begin{align}
    &\text{In NM(SC)}(x>d_F):~~~~~{\tilde E}_z(x)={\cal E}_{s} e^{ -(x-d_F)/\delta(\lambda)},\nonumber\\
    &\text{In NM(SC)}(x<-d_F):~~~~{\tilde E}_z(x)=-{\cal E}_{s} e^{ (x+d_F)/\delta(\lambda)},
\end{align}
where ${\cal E}_s=-\omega\mu_0{\cal M}_y\sinh(i k_0 d_F)/(k_0\cosh(ik_0 d_F)-k_{n(s)}\sinh(i k_0 d_F))$ is the amplitude of the electric field at the interfaces.

The electric fields inside the NM (SC) induce a normal or superconducting current with ${\bf J}_{n(s)}=\sigma_{c(s)}{\bf E}$, which is opposite in the upper and bottom NM (SC). As illustrated in Fig.~\ref{fig:3}, these currents, in turn, generate an Oersted field  $H_y$ inside the FI 
\begin{align}
    H_y&(|x|<d_F)\nonumber\\
    &=\left(\frac{k_0\cosh(ik_0 x)}{k_0\cosh(ik_0 d_F)-k_{n(s)}\sinh(ik_0 d_F)}-1\right)M_y\nonumber\\
    &\approx -\frac{d_F}{d_F+\delta(\lambda)}M_y\label{hy}
\end{align}
noting $k_0d_F\ll 1$, 
where $\delta$ and $\lambda$ represent the penetration depth of microwaves in the NM and SC, respectively. From Eq.~(\ref{hy}), we find such magnetic field is proportional to the magnetization $M_y$, which acts as the effective demagnetization field with the demagnetization factor $N_{yy}=d_F/(d_F+\delta(\lambda))$ given in the main text.

\section{Swihart mode and its quantization}
\label{appendix_Swihart}

In this appendix, we first review the electromagnetic modes in the SC$|$I$|$SC heterostructure. The thickness of the middle non-magnetic insulator is $2d_I$. Swihart found that the superconductor strongly modulates the distribution of the electromagnetic field, significantly slowing down the speed of electromagnetic waves, which is now known as the Swihart mode~\cite{swihart}.

According to Maxwell's equations~\cite{Jackson}, in different regions, the electric field obeys 
\begin{align}
\text{in I}:& ~~~\nabla^2 \mathbf{E}({\bf r}, t)+k^2_0 \mathbf{E}({\bf r}, t)
=0,\nonumber\\
\text{in SC}:& ~~~\nabla^2 \mathbf{E}({\bf r}, t)+k^2_s \mathbf{E}({\bf r}, t)
=0,
\label{waveequ}
\end{align}
where $k_0=\omega\sqrt{\mu_0\epsilon_{\rm I}}$. Our system is isotropic in the $y$-$z$ plane, so we consider the electromagnetic waves propagating along the $\bf \hat{z}$-direction with wavevector $k$ without losing generality. We assume the formal solution ${\bf E}(x,z,t)=\tilde{\bf E}(x)e^{ikz-i\omega t}+{\rm H.c.}$, where the amplitudes
\begin{align}
    &\text{in I}:~~{\tilde {\bf E}}(x)={\pmb{\cal E}}_{0}e^{i {\cal A}_k x}+{\pmb{\cal E}}_{0}'e^{-i {\cal A}_k x},\nonumber\\
    &\text{in SC~``1"}~(x>d_I):~~{\tilde {\bf E}}(x)={\pmb{\cal E}}_{1}e^{i {\cal B}_k x},\nonumber\\
  &\text{in SC~``2"}~(x<-d_I):~~{\tilde {\bf E}}(x)={\pmb{\cal E}}_{2}e^{-i {\cal B}_k x},\nonumber
 \end{align}
where ${\cal A}_k=\sqrt{k_0^2-k^2}$ and ${\cal B}_k=\sqrt{\omega^2\mu_0\epsilon_0-1/\lambda^2-k^2}$. 
In the non-magnetic insulator, the magnetic field is governed by Faraday's Law, i.e., $i\mu_0\omega {\bf H}({\bf r},t)=\nabla\times {\bf E}({\bf r},t)$, leading to
     \begin{align}
     &H_x=1/(i \omega\mu_0)(\partial_y E_z-\partial_z E_y),\nonumber\\
     &H_y=1/(i \omega \mu_0)(\partial_z E_x-\partial_x E_z),\nonumber\\
     &H_z=1/(i \omega\mu_0)(\partial_x E_y-\partial_y E_x).
     \end{align}
     
The boundary conditions are already described in the main text, with which we find the amplitudes $\{{\cal E}_{0x},  {\cal E}_{0x}',    {\cal E}_{0y},  {\cal E}_{0y}'\}$ obey the matrix equation
\begin{widetext}
\begin{equation}
    \left(\begin{matrix}
         (R_k-1)e^{i{\cal A}_k d_I} & (R_k+1)e^{-i{\cal A}_k d_I} & 0 & 0 \\
         (R_k+1)e^{-i{\cal A}_k d_I} & (R_k-1)e^{i{\cal A}_k d_I} & 0 & 0 \\
         0 & 0 & ({\cal A}_k-{\cal B}_k)e^{i{\cal A}_k d_I} & -({\cal A}_k+{\cal B}_k)e^{-i{\cal A}_k d_I} \\
         0 & 0 & ({\cal A}_k+{\cal B}_k)e^{-i{\cal A}_kd_I} & -({\cal A}_k-{\cal B}_k)e^{i{\cal A}_kd_I} \\
    \end{matrix}\right)\left(
    \begin{matrix}
        {\cal E}_{0x}\\
        {\cal E}_{0x}'\\
        {\cal E}_{0y}\\
        {\cal E}_{0y}'
    \end{matrix}
    \right)=0.
\end{equation}
\end{widetext}
This implies that the $x$- and $y$-components of the electric fields decouple, forming two distinct modes. 
The non-zero solutions require the determinant of the coefficient matrix to equal zero.

The first mode is the TE mode with $E_x=E_z=H_y=0$. It has three components $\{H_z,H_x,E_y\}$. From the secular equation 
\begin{align}
\left| \begin{matrix}
({\cal A}_k-{\cal B}_k)e^{i{\cal A}_k d_I} & -({\cal A}_k+{\cal B}_k)e^{-i{\cal A}_k d_I} \\
({\cal A}_k+{\cal B}_k)e^{-i{\cal A}_kd_I} & -({\cal A}_k-{\cal B}_k)e^{i{\cal A}_kd_I} \\
\end{matrix} \right|=0,
\nonumber
\end{align}
we find its dispersion relation 
\begin{equation}
    \omega=\frac{1}{\sqrt{\mu_0\epsilon_{\rm I}}}\sqrt{k^2+\frac{1}{2}\left(\frac{1}{\lambda^2}+\frac{1}{d_F\lambda}\right)}.
\end{equation}
which holds a large cut-off frequency and hence mismatches with the Kittel frequency in the magnetic film.

The second one is the TM mode with $E_y=H_x=H_z=0$. This mode has three components $\{E_z,E_x,H_y\}$. From the secular equation
\begin{align}
    \left|
    \begin{matrix}
    (R_k-1)e^{i{\cal A}_k d_I} & (R_k+1)e^{-i{\cal A}_k d_I}  \\
    (R_k+1)e^{-i{\cal A}_k d_I} & (R_k-1)e^{i{\cal A}_k d_I} 
    \end{matrix}\right|=0, 
    \nonumber
\end{align}
we find the Swihart-mode frequency~\cite{swihart}  
\begin{equation}
\Omega_s=\sqrt{\frac{d_I}{d_I+\lambda}}\frac{1}{\sqrt{\mu_0\epsilon_{\rm I}}}|{\bf k}|.
\label{Swihart_mode}
\end{equation}

Next, we quantize the Swihart mode. The Hamiltonian associated with the Swihart mode of the SC$|$I$|$SC Josephson junction includes the energy of electromagnetic field as well as the supercurrents~\cite{Superfluids}:
\begin{equation}
    \hat{H}_{\rm Sw}=\int{\rm d}{\bf r}\left(\frac{\epsilon_r}{2}\hat{\bf E}^2+\frac{\mu_0}{2}\hat{\bf H}^2+\frac{\mu_0}{2}\lambda^2\hat{\bf J}_{\rm s}^2\right),
    \label{HSw}
\end{equation}
where ${\bf J}_{\rm s}$ is the super-current flowing in the superconductor. We expand the electromagnetic field and supercurrent in terms of the photon operators $\hat{p}_k$ as 
\begin{align}
    \hat{\bf E}&=\int^{+\infty}_{-\infty}\frac{{\rm d}k}{\sqrt{2\pi}}\left(\tilde{\bf E}_{k}(x)e^{ikz}\hat{p}_k+{\rm H.c.}\right),\nonumber\\
    \hat{\bf H}&=\int^{+\infty}_{-\infty}\frac{{\rm d}k}{\sqrt{2\pi}}\left(\tilde{\bf H}_{k}(x)e^{ikz}\hat{p}_k+{\rm H.c.}\right),\nonumber\\
    \hat{\bf J}_s&=\int^{+\infty}_{-\infty}\frac{{\rm d}k}{\sqrt{2\pi}}\left(\tilde{\bf J}_{s,k}(x)e^{ikz}\hat{p}_k+{\rm H.c.}\right),
\end{align}
where $\tilde{\bf E}_k(x)$, $\tilde{\bf H}_k(x)$, and $\tilde{\bf J}_{s,k}(x)$ are the amplitudes. The photon creation and annihilation operators obey the commutation relation $[\hat{p}_k, \hat{p}^{\dagger}_{k'} ]=i\hbar\delta(k-k')$.

When $-d_I<x<d_I$, the electromagnetic field is derived from Eq.~\eqref{waveequ} and the boundary conditions as
\begin{align}
    \tilde{E}_{x,k}(x)&={\cal E}_{0x}e^{i{\cal A}_k x}+{\cal E}_{0x}'e^{-i{\cal A}_k x},\nonumber\\
    \tilde{E}_{z,k}(x)&=\frac{{\cal A}_k}{k}({\cal E}_{0x}e^{i{\cal A}_k x}-{\cal E}_{0x}'e^{-i{\cal A}_k x}),\nonumber\\
    \tilde{H}_{y,k}(x)&=\frac{{\omega \epsilon_{\rm I} }}{k}({\cal E}_{0x}e^{i{\cal A}_k x}+{\cal E}_{0x}'e^{-i{\cal A}_k x}),\nonumber\\
    \tilde{J}_{s,k}(x)&=0.
    \label{distribution1}
\end{align}
They are incorporated into the electromagnetic field at $x>d_I$ and $x<-d_I$ after applying the continuity conditions for the $E_z$ and $H_y$ at the interfaces:
    \begin{align}
        \tilde{E}_{x,k}(x)&=\frac{\epsilon_{\rm I}}{\epsilon_0+i\sigma_s/\omega}\tilde{E}_{x,k}(\pm d_I^{\mp})e^{\pm i{\cal B}_k (x\mp d_I)},\nonumber\\
        \tilde{E}_{z,k}(x)&=\tilde{E}_{z,k}(\pm d_I)e^{\pm i{\cal B}_k (x\mp d_I)},\nonumber\\
        \tilde{H}_{y,k}(x)&=\tilde{H}_{y,k}(\pm d_I)e^{\pm i{\cal B}_k (x\mp d_I)},\nonumber\\
        \tilde{J}_{s,k}(x)&=\sigma_s\tilde{E}_{z,k}(x)=\frac{i}{\omega\mu_0\lambda^2}\tilde{E}_{z,k}(\pm d_I)e^{\pm i{\cal B}_k (x\mp d_I)}.\label{distribution2}
    \end{align}
Here with the dispersion of the Swihart mode \eqref{Swihart_mode}
\begin{align}
     {\cal A}_k&=\sqrt{k_0^2-k^2}=ik\sqrt{\frac{\lambda}{d_I+\lambda}},\nonumber\\
     {\cal B}_k&=\sqrt{k_s^2-k^2}=i\sqrt{\frac{1}{\lambda^2}+\frac{\lambda}{d_I+\lambda}k^2}.\nonumber
\end{align}
When the wavelength of the photon is much larger than the scale of the system, i.e., $k\ll 1/\lambda$ and $k\ll 1/d_I$, ${\sin 2{\cal A}_k d_I}/({2{\cal A}_k})\approx d_I$, $\cos 2{\cal A}_k d_I \approx 1$, and 
\begin{align}
    {\cal B}_{k}=i\sqrt{\frac{1}{\lambda^2}+\frac{\lambda}{d_{I}+\lambda}k^2}\approx \frac{i}{\lambda}.\nonumber
\end{align}
Furthermore, the amplitudes
\begin{align}
    {\cal E}_{0x}'&=\frac{({\cal B}_k+k^2){\cal A}_k-({\cal A}_k+k^2){\cal B}_k}{({\cal B}_k^2+k^2){\cal A}_k+({\cal A}_k^2+k^2){\cal B}_k}e^{2i{\cal A}_kd_I}{\cal E}_{0x}\nonumber\\
    &\approx \frac{1+\frac{d_I}{d_I+\lambda}k\lambda}{1-\frac{d_I}{d_I+\lambda}k\lambda}{\cal E}_{0x}\approx{\cal E}_{0x}.\nonumber
\end{align}

We normalize ${\cal E}_{0x}$ according to 
\begin{align}
    &\frac{1}{2}\int {\rm d}x [{\epsilon_r}\left(\tilde{E}_{x,k}\tilde{E}^*_{x,k}+\tilde{E}_{z,k}\tilde{E}^*_{z,k}\right)\nonumber\\
    &+{\mu_0}\tilde{H}_{y,k}\tilde{H}^*_{y,k}+{\mu_0}\lambda^2\tilde{J}_{s,k}\tilde{J}_{s,k}^*]
    \nonumber\\
    &\approx 4\epsilon_{\rm I}{\cal E}_{0x}{\cal E}_{0x}^{*}d_I\left(1+{k_0^2}/{k^2}\right)+2\epsilon_{\rm I}{\cal E}_{0x}{\cal E}_{0x}^{*} \lambda\nonumber\\
    &\times\left(\frac{k_0^2}{k^2}+\frac{1}{2k^2d_I\lambda}\left(-\frac{1}{2}4{\cal A}_k^2 d_I^2\right)\right)\nonumber\\
    &=8\epsilon_{\rm I}{\cal E}_{0x}{\cal E}_{0x}^{*}d_I={\hbar\Omega_s}/{2}.
    \label{Hkk}
\end{align}
Then with relation
\begin{align}
    &\frac{1}{2}\int {\rm d}x [{\epsilon_r}\left(\tilde{E}_{x,k}^{(*)}\tilde{E}_{x,-k}^{(*)}+\tilde{E}_{z,k}^{(*)}\tilde{E}_{z,-k}^{(*)}\right)\nonumber\\
    &+{\mu_0}\tilde{H}_{y,k}^{(*)}\tilde{H}_{y,-k}^{(*)}+{\mu_0}\lambda^2\tilde{J}_{s,k}^{(*)}\tilde{J}_{s,-k}^{(*)}]\nonumber\\
    &=4\epsilon_{\rm I}{\cal E}_{0x}^{(*)}{\cal E}_{0x}^{(*)}d_I\left(1-\frac{k_0^2}{k^2}\left(1+\frac{\lambda}{d_I}\right)\right)=0
    \label{Hk-k}
\end{align}
and by substitution of Eqs.~\eqref{Hkk} and \eqref{Hk-k} into Eq.~\eqref{HSw}, we arrive at the Hamiltonian of Swihart mode   
$\hat{H}_{\rm Sw}=(1/2)\int {\rm d}k \hbar\Omega_s\left(\hat{p}_k^{\dagger}\hat{p}_k+\hat{p}_k\hat{p}_k^{\dagger}\right)$.

\end{appendix}

\end{document}